\documentclass[aprilunum,article,pdftex]{Definitions/apunum} 

\firstpage{1} 
\pubvolume{1}
\issuenum{1}
\articlenumber{1}
\pubyear{2026}
\copyrightyear{2026}
\datereceived{2026 Apr 1 } 
\dateaccepted{2026 Apr 1} 
\datepublished{2026 Apr 1} 
\dateretracted{2026 Apr 1} 
\pdfoutput=1 

\usepackage{geometry}
\usepackage{rotating}

\Title{Where to Search For Life: Evidence from narrative sources with established predictive efficacy.}

\TitleCitation{Where to Search for Life}

\Author{Stanway, Elizabeth R. $^{\ast,\dagger}$}

\AuthorNames{Stanway, Elizabeth R}

\address[1]{$^\ast$ \hspace{0.2cm} Centre for Exoplanets \& Habitability and Department of Physics, University of Warwick, UK \footnote{although they may disown me after this}}

\corres{Correspondence: e.r.stanway@warwick.ac.uk}

\firstnote{Please see \textbf{Appendix A} for important clarifications.}  

\abstract{
The search for habitable planets, and even for ``Earth 2.0'', is a major driver in contemporary astronomy. However selecting target fields to prioritise for such searches presents a challenge. Here we establish a statistical analysis of the appearance of constellation names in science fiction magazines of the pulp era, evaluating the most commonly mentioned constellations and thus those which the science fiction community collectively identify as the most likely locations to find life. Given that the predictive power of science fiction is well established, we suggest that these locations might be prioritised by searches for extrasolar biospheres.}


\begin{document}


\section{Introduction}

The search for habitable planets, and even for Earth 2.0, is a major driver in contemporary astrophysics, and connections to exoplanet astronomy are used to motivate (and attract funding to) research on topics which have only the most tenuous relation to astrobiology. A large number of (primarily but not exclusively) space missions have been, or soon will be, launched in the effort to better constrain and interpret the properties and demographics of extra-solar planets. However such studies must inevitably reach a compromise between the size of the survey area examined and the properties of the planets probed. 

Wide-area survey telescopes such as the Transiting Exoplanet Survey Satellite \cite[TESS,][]{2015JATIS...1a4003R} and the upcoming PLAnetary Transits and Oscillations of Stars \cite[PLATO,][]{2025ExA....59...26R} mission search for planet-sized objects by the monitoring of relatively bright stars for the fluctuations in light caused by transiting exoplanets. They survey large areas of the sky, and hence many potential planet hosting suns, albeit with a cadence and sensitivity that varies depending on sky location. 
By contrast, facilities such as the former \textit{Kepler} space telescope \cite{2010Sci...327..977B} and the planned \textit{Habitable Worlds Observatory}, as well as radial velocity surveys, are designed for more sensitive observations targetting fainter stars but have a narrow field of view. They rely either on targetted follow up of known exoplanet systems, or on high cadence monitoring of relatively small target areas.   

Any \textit{a priori} information regarding regions which might be likely to contain habitable environments can, in principle, be used to inform survey design, particularly for small-field or focussed surveys. Given the investment  being focussed on exoplanet science, it is clear that we should exploit all possible sources of such prior information - including those which have traditionally been neglected by the astronomical community. 

\bigskip

One such source is speculative fiction The predictive power of science fiction is widely recognised \cite[see e.g.][]{brinkhof_2023_science, mcnamara_2019_7,handwerk_2016_the,murdock_2010_6,GerroldPredictions}. The genre is known to have made early predictions of solar sails \cite[e.g.][]{VerneLaLune}, exploitation of solar power \cite[e.g.][]{GernsbackRalph,LeinsterPowerPlanet}, in vitro fertilisation and genetic engineering \cite[e.g.][]{HaldaneDaedalus,Huxley1932}, domestic robots \cite[e.g.][]{Capek_RUR,Bradburysoftrains}, reality television \cite[e.g.][]{Kneale_SexOlympics} and global telecommunications \cite[e.g.][]{ForsterMachineStops} amongst many other examples. The magazine \textit{Astounding Science Fiction}, its editor and authors were famously investigated by US wartime intelligence services when some of its stories \cite{heinleinsolution,CartmillDeadline}  proved uncomfortably close to the contemporary and top secret Manhattan Project in terms of their technology \cite{BergerAstounding}. 

As a community, both science fiction (SF) readers and writers have a reputation for scientific literacy that exceeds that of the general public. Some of the earliest science fiction was published in journals aimed at a technical readership (e.g. \textit{Ralph 124C 41+} by Hugo Gernsback was serialised in \textit{Modern Electrics} in 1911-1912 \cite{GernsbackRalph}). When SF developed its own periodicals, these routinely accompanied fiction with factual information -- much of it on scientific matters of relevance to the narratives. Prolific contributors of scientific articles included science writer Willy Ley (who wrote a regular feature called \textit{For Your Information} in \textit{Galaxy Science Fiction} magazine from 1952 to 1969), astronomer Robert S.~Richardson (who wrote regularly for \textit{Astounding} amongst other publications), biochemist Isaac Asimov (who wrote regular science columns for \textit{Fantasy \& Science Fiction} magazine) and psychologist Jerry Pournelle (who wrote regular factual columns for \textit{Galaxy} and later \textit{Asimov's}). 

A profile of US science fiction readers published in 1954 suggested that the readership were, on average, more highly educated than the general public and that around 40\% were employed in technical, engineering or scientific research roles \cite{doi:10.1111/j.1467-954X.1954.tb00972.x}. A meta-analysis of subsequent surveys by science fiction journals, undertaken in 1977, showed similar results, with between 20 and 40 percent of readers typically employed in science and technical roles \cite{10.1525/sfs.4.3.0232}, although the author noted a decline into the 1970s as social sciences became more strongly represented. More recent studies of science fiction readers have seen similar trends, with a 2018 study suggesting that almost 70\% of SF readers feel it helps them relate to science \cite{doi:10.1177/2158244018780946}. 

Supporting these statistics are the numerous anecdotal accounts from science fiction writers of interactions with readers who have noticed or engaged with the scientific content of their narratives, and in particular errors therein. The mutual understanding of this co-creative pressure was articulated by hard science fiction writer Hal Clement in a factual science article \cite{whirligig} accompanying publication of his novel \textit{A Mission of Gravity} in 1953:

\begin{quote}
\textit{
I've been playing the game since I was a child, so the rules must be quite simple. They are; for the reader of a science-fiction story, they consist of finding as many as possible of the author's statements or implications which conflict with the facts as science currently understands them. For the author, the rule is to make as few such slips as he possibly can.} 
\end{quote}

While narrative necessity may override scientific precision where required, such `games' lead to a pressure on science fiction writers (many of whom were themselves scientifically trained, including examples mentioned above) to present an accurate representation of contemporary scientific knowledge in their stories. 

\bigskip

Science fiction narratives, in the form of novels, novellas and short stories, have traditionally been published in a range of periodical magazines. The most widely circulated and popular of these were initially founded in the 1920s and 30s, and were typically associated with poor quality paper in order to ensure low cover prices. As a result, they became known as the science fiction \textit{pulp} magazines \cite{sfe_entry_pulp}, a name which stuck even as the quality of the paper improved over time and magazines became more commonly published in the \textit{digest} format. The peak era of the pulps was between the 1940s and 1960s, when such magazines were the primary forum for publication of leading science fiction authors such as Isaac Asimov, Arthur C Clarke, Robert A Heinlein and others. While the pulp magazines declined in circulation with the rise of domestic television, a number continued in press into the last quarter of the twentieth century, and a handful continue to be published today.

The rise of online archives, and in particular the Internet Archive\footnote{https://archive.org/}, has made large collections of pulp SF magazines available for study, with many fully indexed  so the magazine text is searchable. Here we present a statistical analysis of the appearance of constellation names in science fiction magazines of the pulp era, identifying the most commonly mentioned constellations and thus establishing those which the science fiction community collectively identify as the most likely locations to find life. Leveraging the predictive power of speculative fiction, we discuss possible implications for upcoming exoplanet astronomy survey prioritisation.

\section{Methodology}

\begin{figure}
\hspace{-0.1cm}
\includegraphics[height=0.65\textwidth]{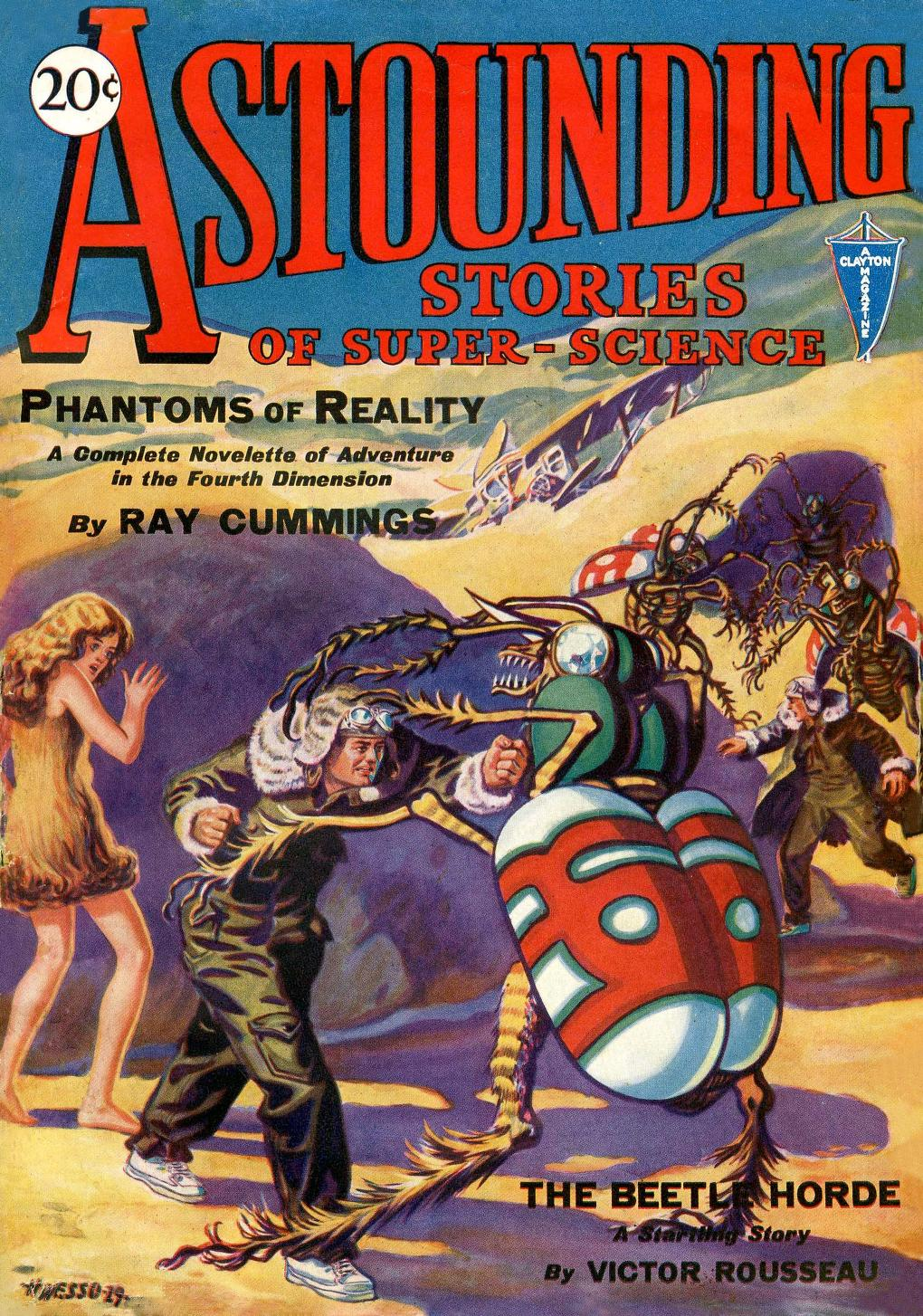}
\includegraphics[height=0.65\textwidth]{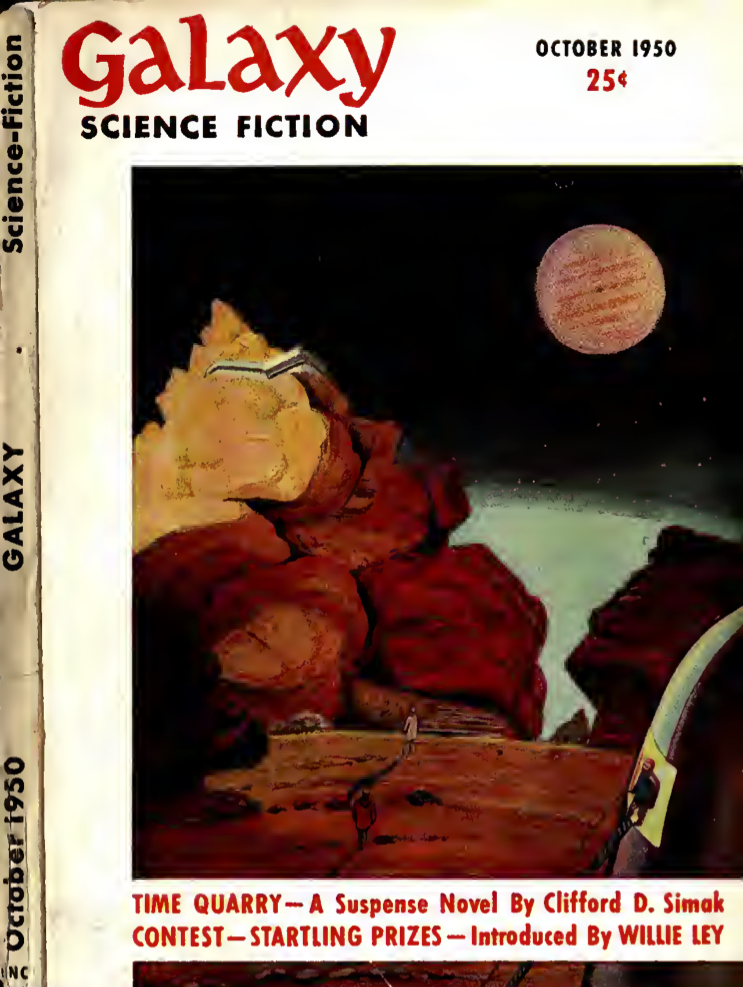}
\caption{Pulp magazine covers, showing the evolution from sensationalism towards more carefully considered illustrations of astronomical scenes. (Left) Cover of Astounding Stories of Super-Science, volume 1, issue 1 (1930). Illustration by H W Wessolowski, accompanying The Beetle Horde by Victor Rousseau. (Right) Cover of Galaxy Science Fiction magazine, volume 1, issue 1 (1950). Illustration by David Stone, accompanying serialised novel Time Quarry by Clifford D Simak. Image source: archive.org. Images out of copyright.}\label{fig:galaxycover}
\end{figure}
\subsection{Literature Sources}

We make use of the extensive pulp magazine archives available on the donation-supported open-access database \textit{archive.org}. A large fraction of these have undergone full indexing such that searches for matches in the inside text of individual magazines are possible (albeit subject to the limitations of optical character recognition). We consider six such indexed collections:

\smallskip

\textit{Amazing Stories} was one of the leading pulp science fiction magazines. First issued in April 1926, under the editorship of Hugo Gernsback, it continued to be published monthly until the end of 1963, and thereafter bi-monthly until the end of 1976 and under a range of other frequencies, from monthly to quarterly into the 1990s.  
The archive.org collection \cite{amazingstories} comprises 779 separate files, the majority each corresponding to a different issue of \textit{Amazing Stories}, although there is a small amount of duplication resulting from a range of contributors, who use different file naming conventions. Visual inspection of search results suggest about 10\% of results are typically duplications of the same issue and these are removed from number counts where identified.

\smallskip

\textit{Isaac Asimov's Science Fiction Magazine} (known as \textit{IAsfm} and later simply as \textit{Asimov's}) was founded in 1977, with a remit to lean towards short-form fiction and towards the harder end of the science fiction spectrum. Its title was shortened to Asimov's Science Fiction in 1992. A variety of editors over the years have included the long-serving Gardner Dozois (1985-2004). The magazine continues to be published six times a year to date. 
The collection available on archive.org \cite{asimovs} contains 432 files, with editions of the magazine extending from 1977 to 2015 and very little duplication.

\smallskip

\textit{Astounding Stories of Super Science} was first published in January 1930. It was renamed simply \textit{Astounding Stories} in 1931, and to \textit{Analog Science Fact \& Fiction} in 1960. Its influential editors have included John W.~Campbell Jnr and Ben Bova. Published monthly through the majority of its history, it continues to be issued as \textit{Analog Science Fiction \& Fact} (since 1992) on a bimonthly basis.  
The archive.org collection \cite{astounding} comprises 426 separate files, dating from 1930 to end of the \textit{Astounding Stories} name in December 1959. Again, there is some duplication, including international editions of some issues.  These are removed from search results on inspection, whenever two issues appear in the same month. 

\smallskip

\begin{figure*}
\begin{minipage}{1.15\textwidth}
\hspace*{-4cm}
\includegraphics[width=0.55\textwidth]{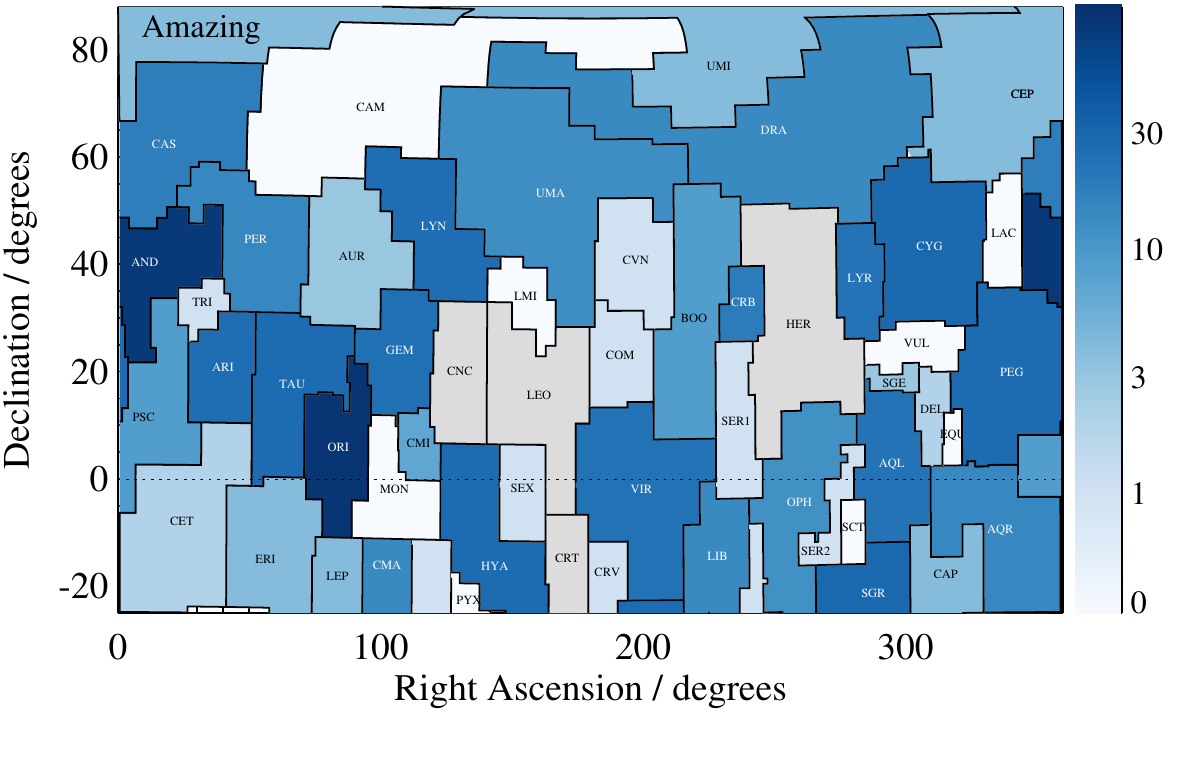}
\includegraphics[width=0.55\textwidth]{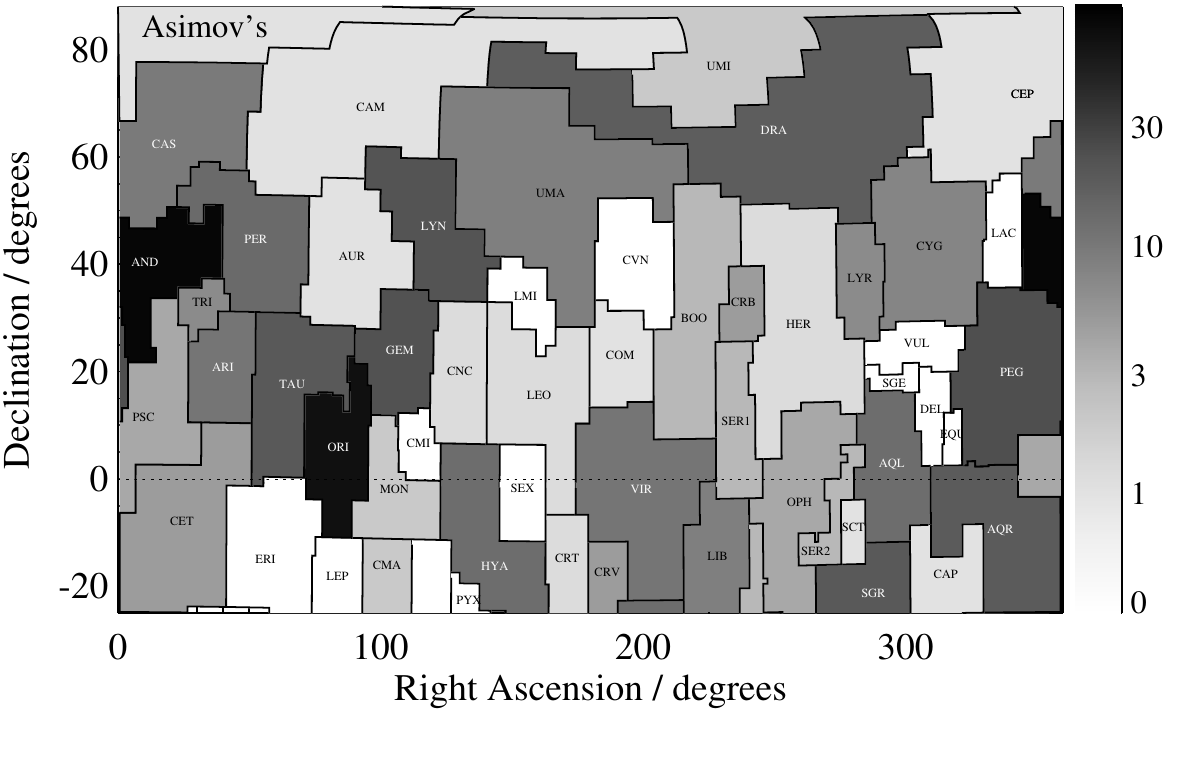}

\hspace*{-4cm}
\includegraphics[width=0.55\textwidth]{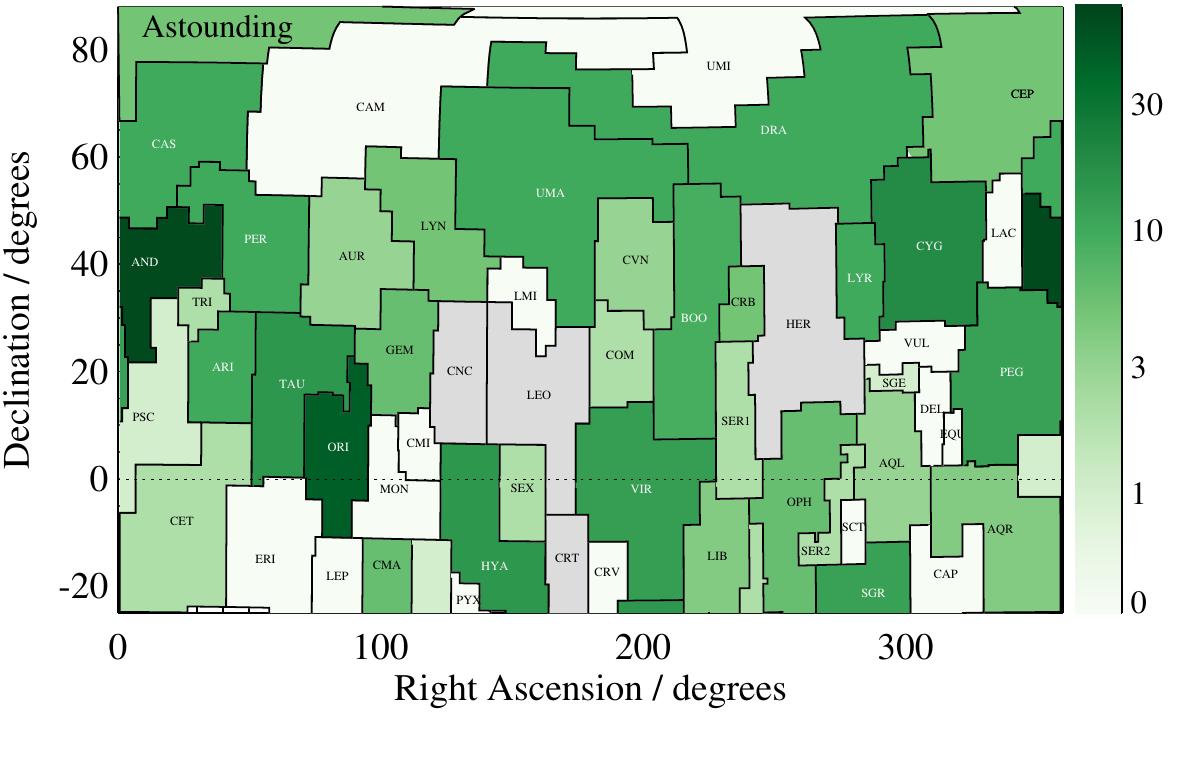}
\includegraphics[width=0.55\textwidth]{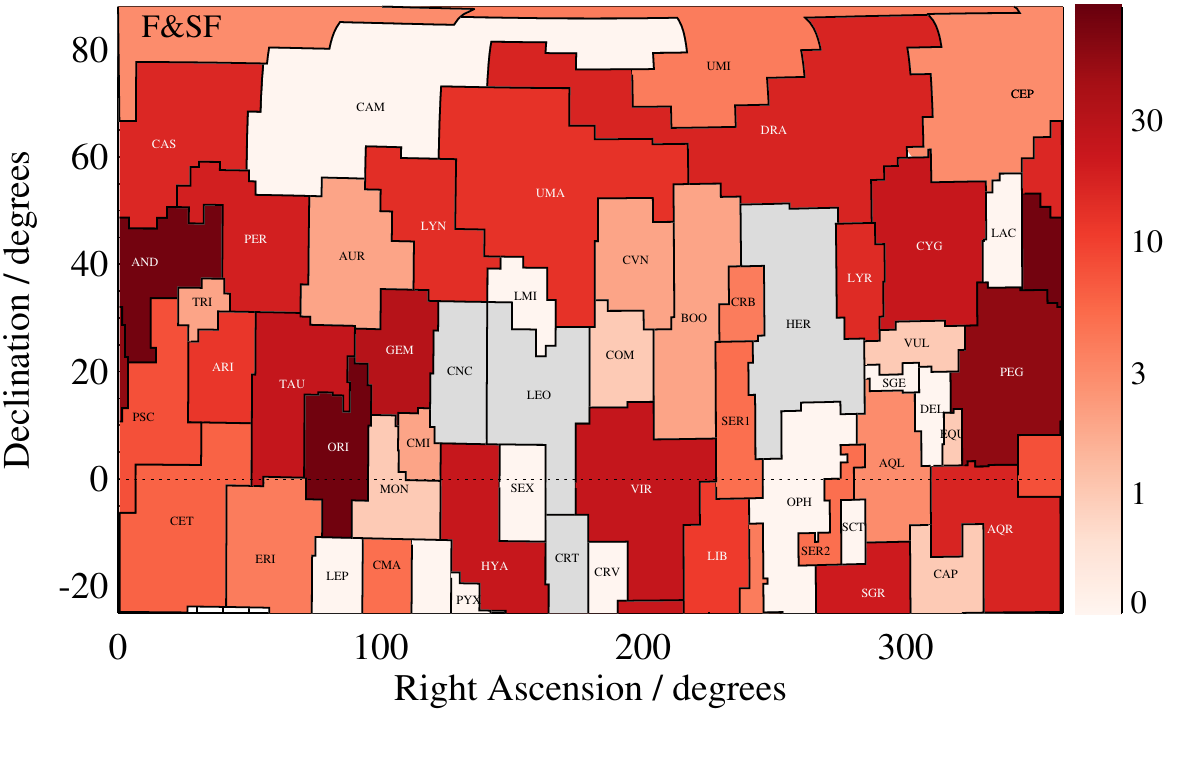}

\hspace*{-4cm}
\includegraphics[width=0.55\textwidth]{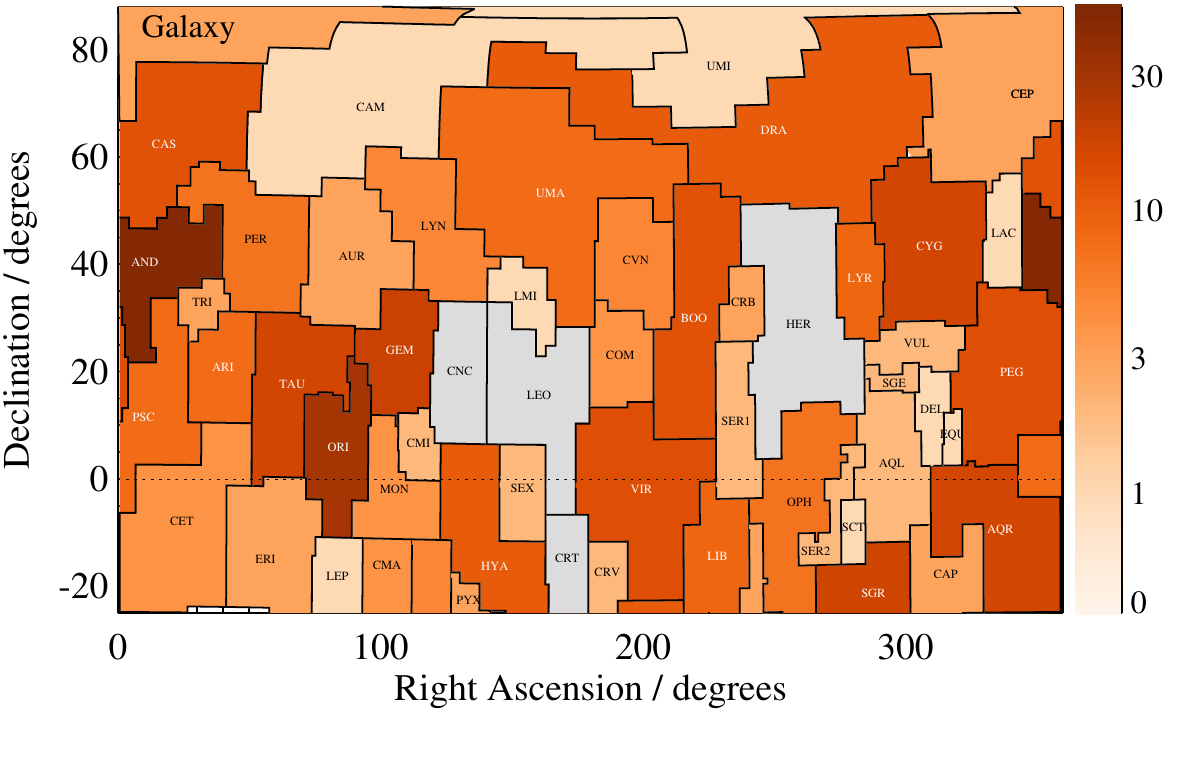}
\includegraphics[width=0.55\textwidth]{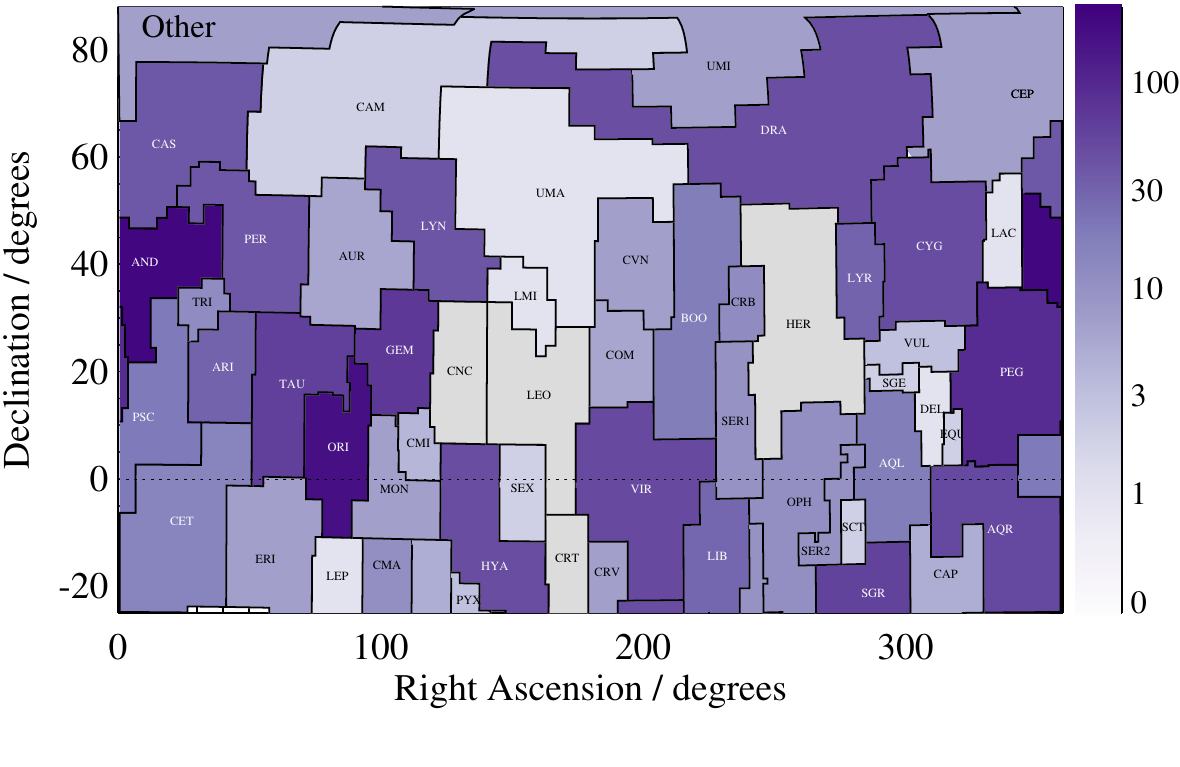}

\end{minipage}
\hspace*{-2cm}
\begin{minipage}{0.99\textwidth}
\caption{References to northern constellations by name, mapped onto equatorial coordinates. The magazine source is given in the top right hand corner of each panel. The celestial equator is marked with a dashed line. More frequently mentioned constellations are indicated by darker shades. Constellations are labelled by acronym as given in Table B1. Data for Crater, Hercules, Cancer and Leo are omitted.}\label{fig:byname}
\end{minipage}
\end{figure*}

\textit{The Magazine of Fantasy \& Science Fiction}, familiarly known as F\&SF, was created in 1949. Initially bimonthly, it shifted to monthly publication and then back to bimonthly, at which frequency it continues until the current time. As the name suggests, its remit includes fantasy (i.e. scenarios based on imagination) as well as science fiction (with scenarios based on extrapolation and technology). 
The archive.org collection \cite{fandsf} comprises 607 individual files, including editions of the magazine dating from 1949 to 2007. There is a very small amount of duplication, in the form of international editions and anthologies, but this is at the level of less than one percent. No removal of search result duplications was required.

\smallskip

\textit{Galaxy Science Fiction} was published monthly between 1950 and 1980. Notable editors included H L Gold, Frederik Pohl and Judy-Lynn del Rey. Galaxy launched with an editorial stating its intent to deliver a more adult and respectable form of science fiction, starting with commissioning more refined and scientifically motivated cover art than earlier pulps, as Figure \ref{fig:galaxycover} demonstrates. \textit{Galaxy} is also notable for having contributed many short stories for adaptation by popular NBC radio drama anthology series \textit{Dimension-X} and \textit{X-Minus-One}, which were broadcast in 1950-1951 and between 1955 and 1959 in the United States of America, bringing science fiction to a broader audience than the pulp magazines alone. 
The archive.org collection \cite{galaxy} comprises 446 files spanning the publishing history of the magazine, including 42 novels published as stand-alone books by the publisher. These novels are excluded from searches. About a quarter of the remaining files are duplications of issues, and these are removed from search results.

\smallskip

We also combine search results for three smaller magazines: 


\textit{Worlds of If} (collection: 174 items \cite{ifmagazine}) was first published in March 1952 with a stated intention of focussing on good stories. It peaked in circulation and recognition in the early 1960s, under the editorship of Frederik Pohl, and merged with \textit{Galaxy SF} at the end of 1974. The collection on archive.org has negligible repetition. 


\textit{New Worlds} (collection: 149 items \cite{newworlds}) was a British magazine published regularly between 1939 and the mid-1960s, and intermittently thereafter under a range of funding models, including self-publishing by the editor. Under the leadership of Michael Moorcock, it specialised in more avant-garde and unconventional SF stories.  The archive collection includes editions mainly published between 1950 and 1970, with some duplication and international editions.


\textit{Planet Stories} (collection: 139 items \cite{planetstories}) was a relatively small magazine published between 1939 and 1955. Initially aimed at a younger audience, it occasionally published stories from leading authors. As the name suggests, it specialised in planetary romances - adventure stories set on alien worlds (often but not exclusively in our own Solar System). The collection on archive.org has a relatively high degree of repetition, which is removed from search results by hand.

\subsection{Identification of Astronomical Targets}
As potential targets, we consider the list of constellations defined by the International Astronomical Union. We search for the approved name of the constellation (e.g. Draco or Pegasus) and its genitive form (i.e. draconis or pegasi) since these are often used as components of star names in the constellation.  
We consider these separately, and do not attempt to sum them, since in a number of cases the constellation name and its genitive form will appear in the same text. Each issue (typically monthly) in which the term appears is counted as a hit, and duplicate appearances of the same issue in search results are removed. We also do not attempt to account for multiple appearances in the same issue of a magazine (i.e. appearance in two separate stories appearing the same month would be counted as one hit). Where both US and international magazine editions for the same month are selected, only one is counted.
We do not distinguish between appearance in science fiction stories, editorial or advertising content, or in the science fact articles which routinely appear alongside the fiction, since this is beyond the scope of the current work\footnote{If anyone is interested in employing me to spend the next decade reading the context of each hit to classify it, I would welcome research grant contract offers at the correspondence address.}. The sole exception to this is the December 1963 edition of \textit{Galaxy Science Fiction}, in which a factual article by science populariser Willy Ley names every constellation \cite{ley_Dec1963}. This reference is removed from the collated data. 

Since the majority of writers and readers of the SF pulp magazines considered were WEIRD\footnote{i.e. from Western, Educated, Industrialised, Rich, Democratic countries \cite{Henrich_Heine_Norenzayan_2010}} and, indeed, the majority were American\footnote{The USA currently appears to be trying hard to lose the E, R and D qualifications, but was weird enough during the epoch studied.}, we restrict our analysis to constellations which extend further north than a declination of -30$^\mathrm{o}$. This yields a sample of 54 constellations, but notably excludes the constellations Centaurus (famously home to our stellar neighbours $\alpha$ Centauri and Proxima Centauri) and Piscis Austrinus (less-famously home to another science fictional favourite, Formalhaut). A companion study of the southern hemisphere is deferred to later work.
We note that the official name of the constellations Cancer, Crater, Hercules and Leo present a challenge to interpretation, since the majority of search results do not appear to relate to the constellation. We omit these from analysis, although we retain searches for the genitive names for these constellations.

\begin{figure*}
\begin{minipage}{1.15\textwidth}
\hspace*{-4cm}
\includegraphics[width=0.55\textwidth]{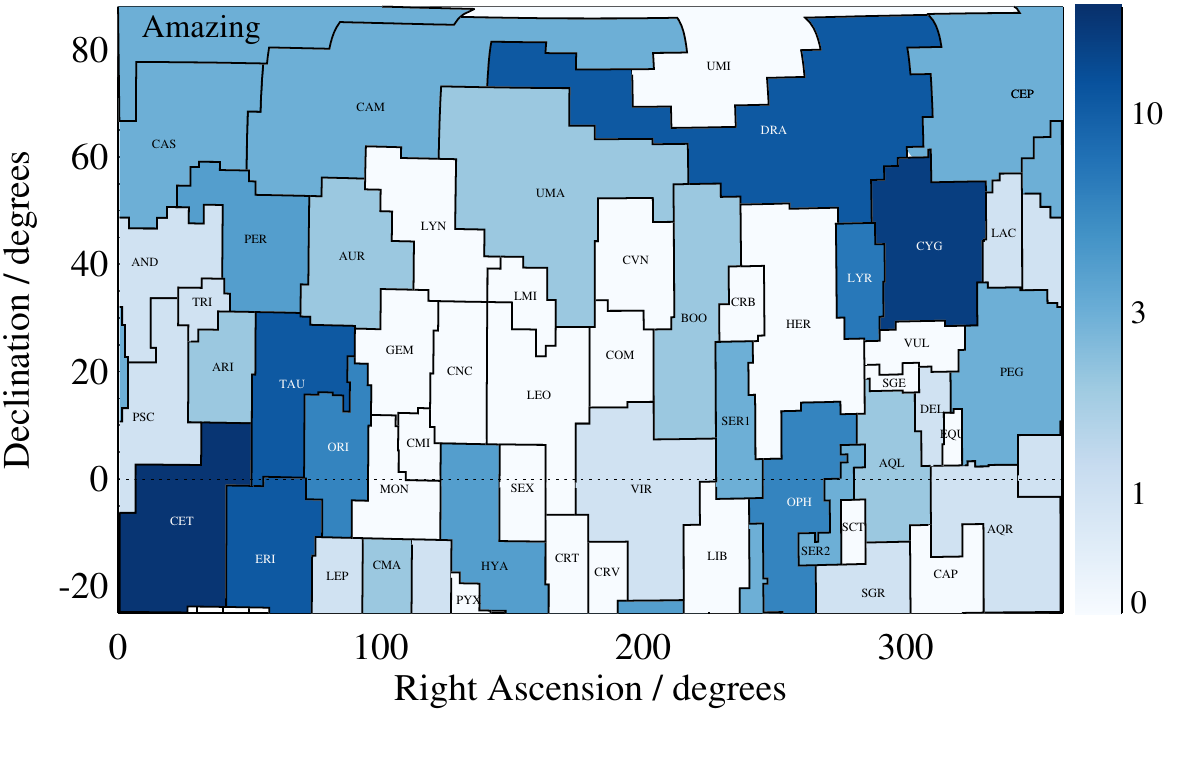}
\includegraphics[width=0.55\textwidth]{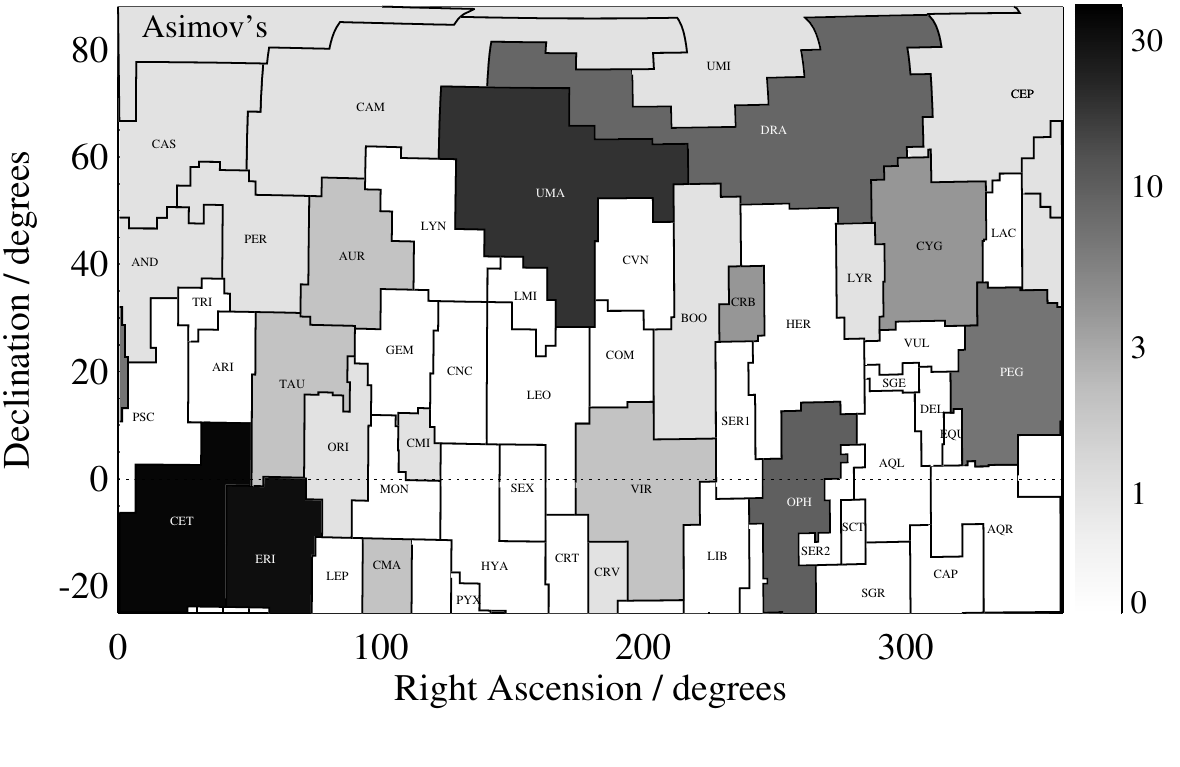}

\hspace*{-4cm}
\includegraphics[width=0.55\textwidth]{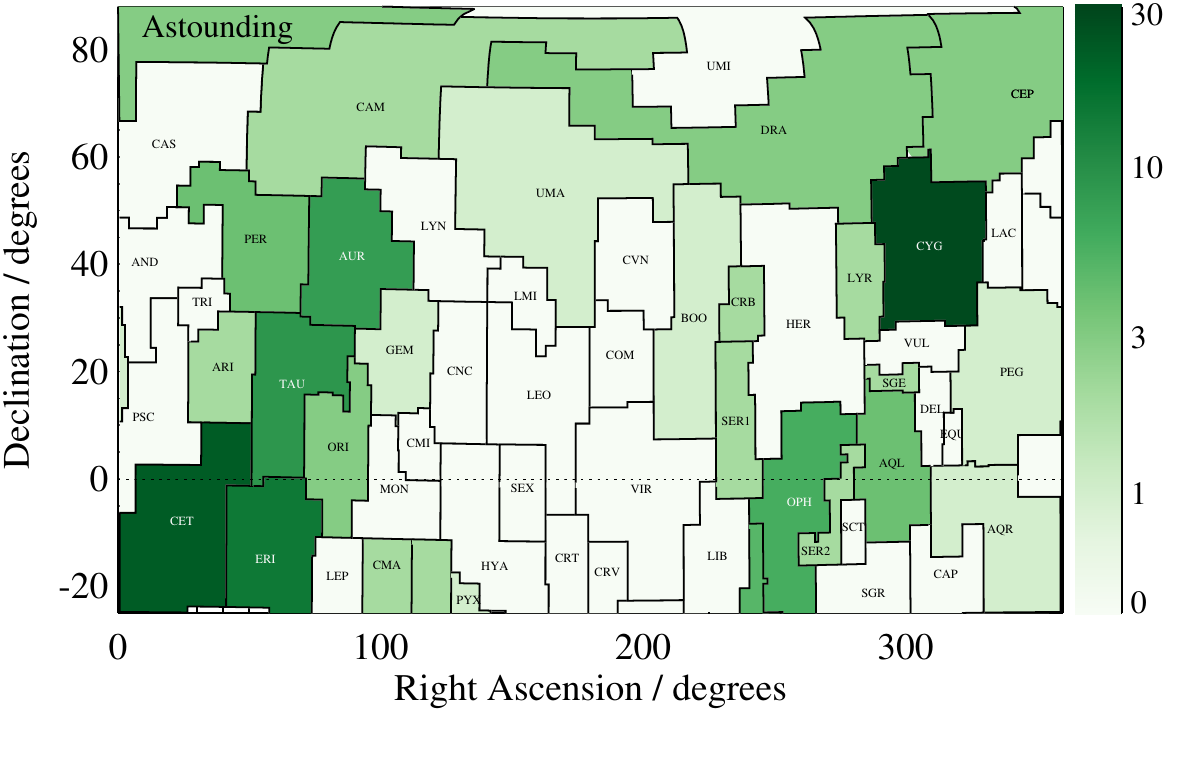}
\includegraphics[width=0.55\textwidth]{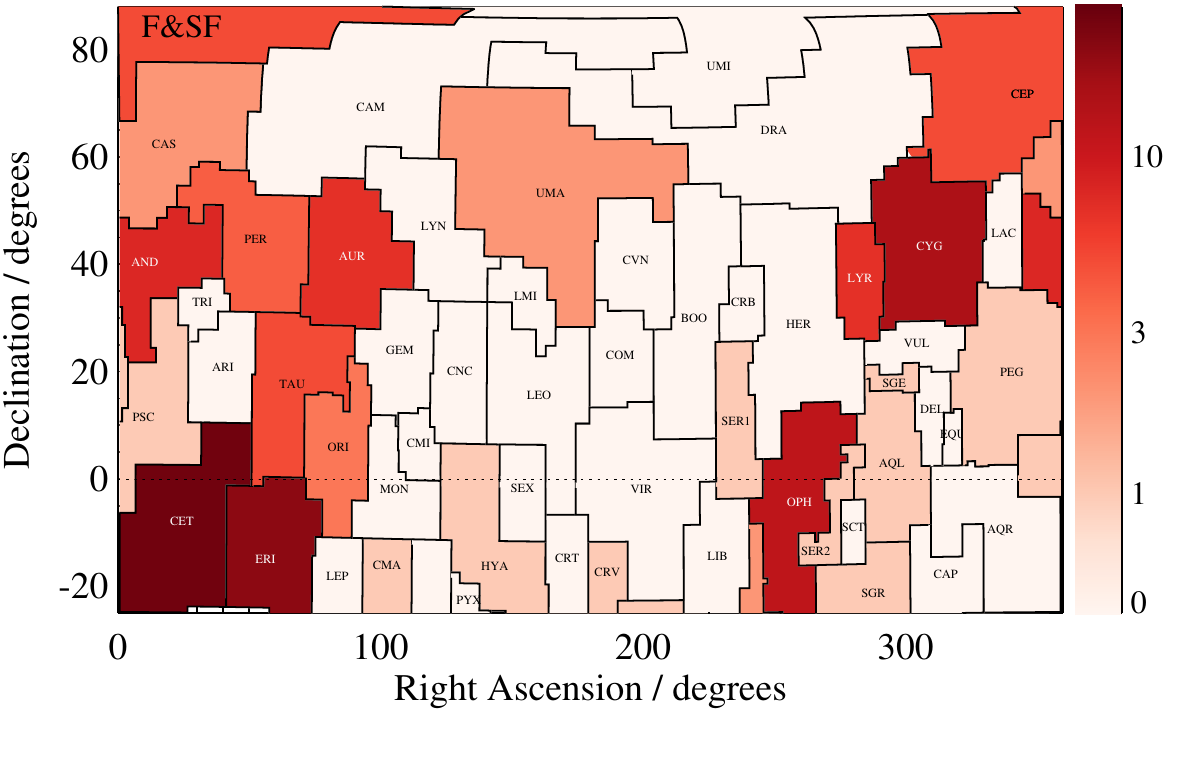}

\hspace*{-4cm}
\includegraphics[width=0.55\textwidth]{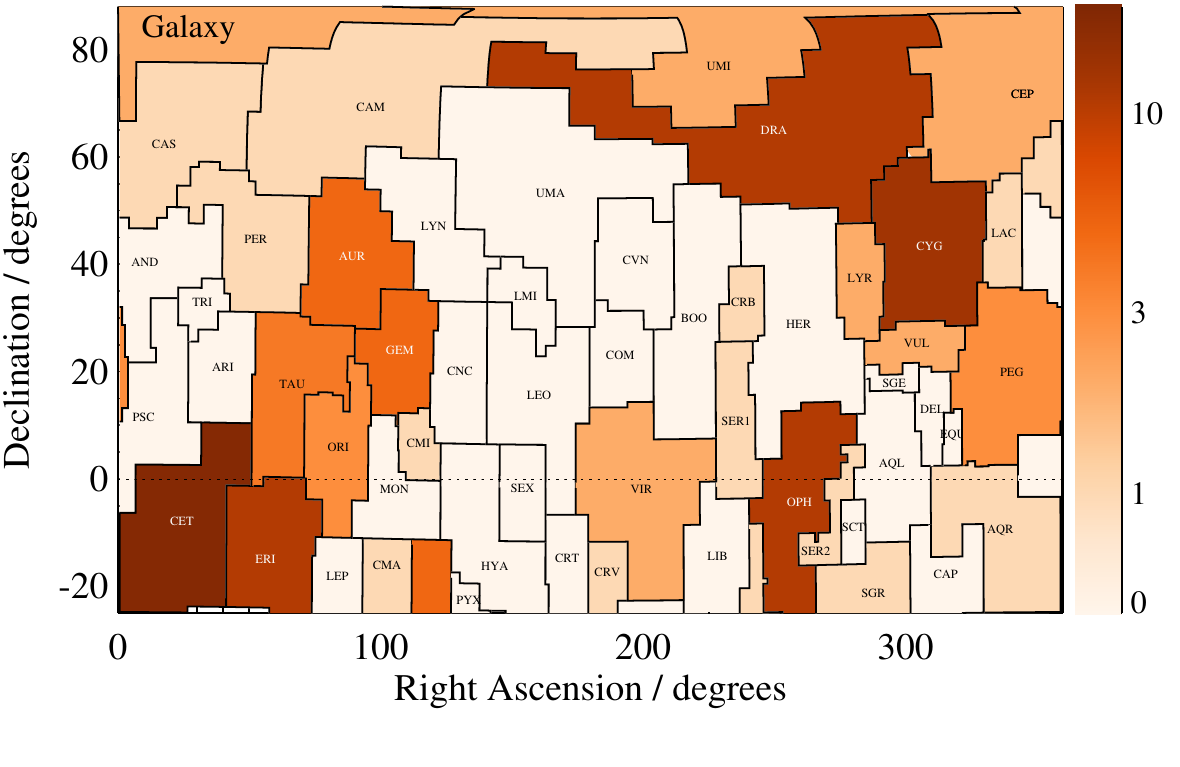}
\includegraphics[width=0.55\textwidth]{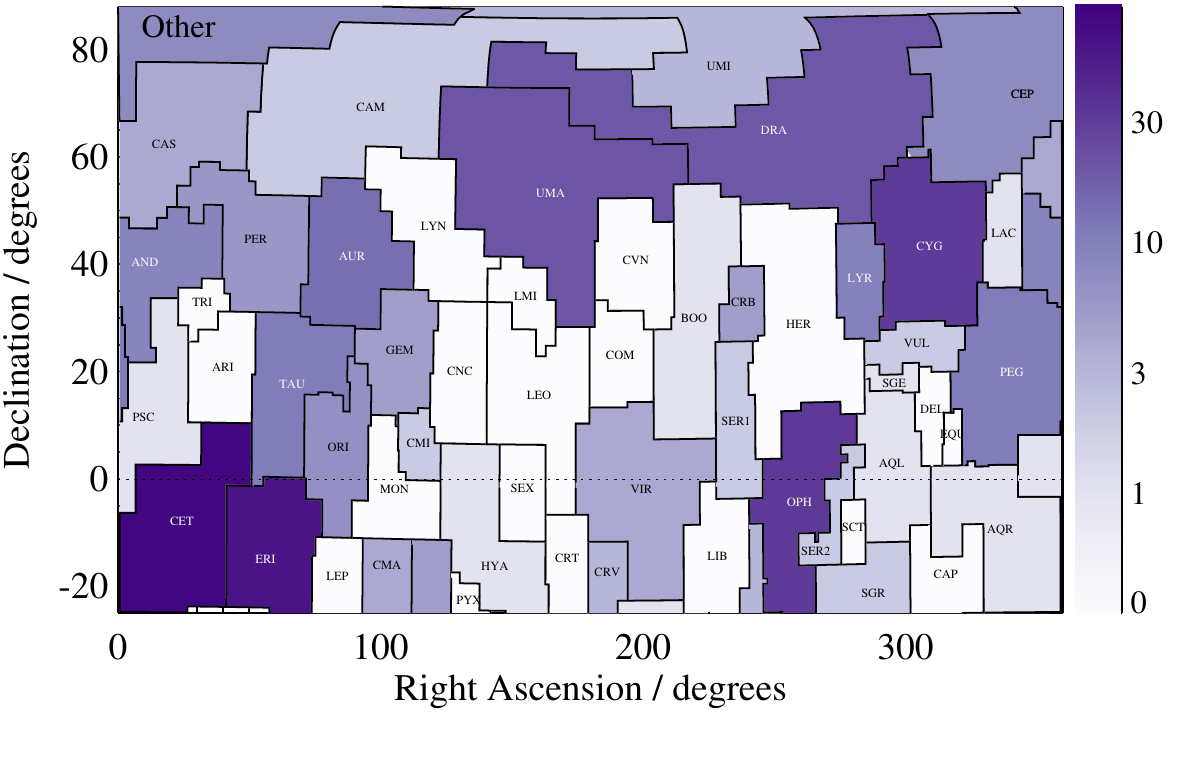}
\end{minipage}
\hspace*{-2cm}
\begin{minipage}{0.99\textwidth}
\caption{ References to constellations by genitive form, as in Figure \ref{fig:byname}. }\label{fig:genitive}
\end{minipage}
\end{figure*}


\section{Results}


\begin{figure*}
\hspace{-1cm}\includegraphics[width=0.98\textwidth]{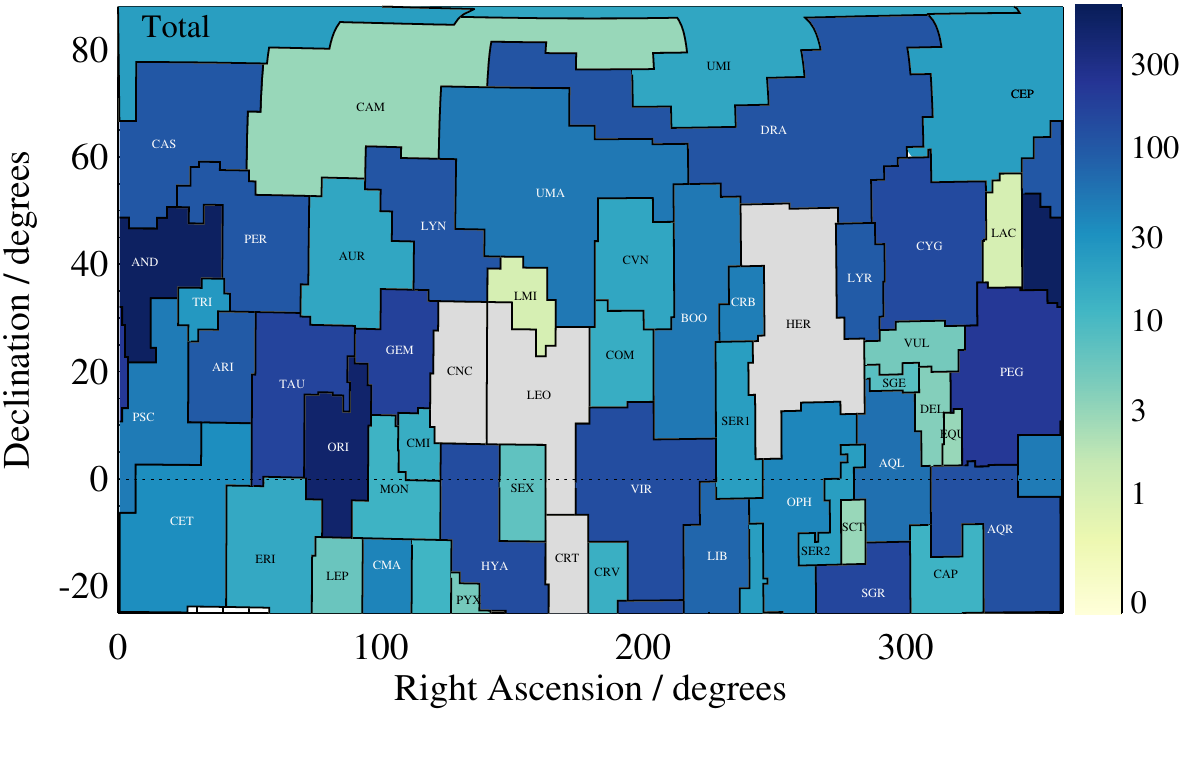}
\caption{Total references to constellations by name. References have been summed across all magazines considered. The four constellations shown in grey have been omitted from searches.}\label{fig:Totalname}
\end{figure*}

\begin{figure*}
\hspace{-1cm}\includegraphics[width=0.98\textwidth]{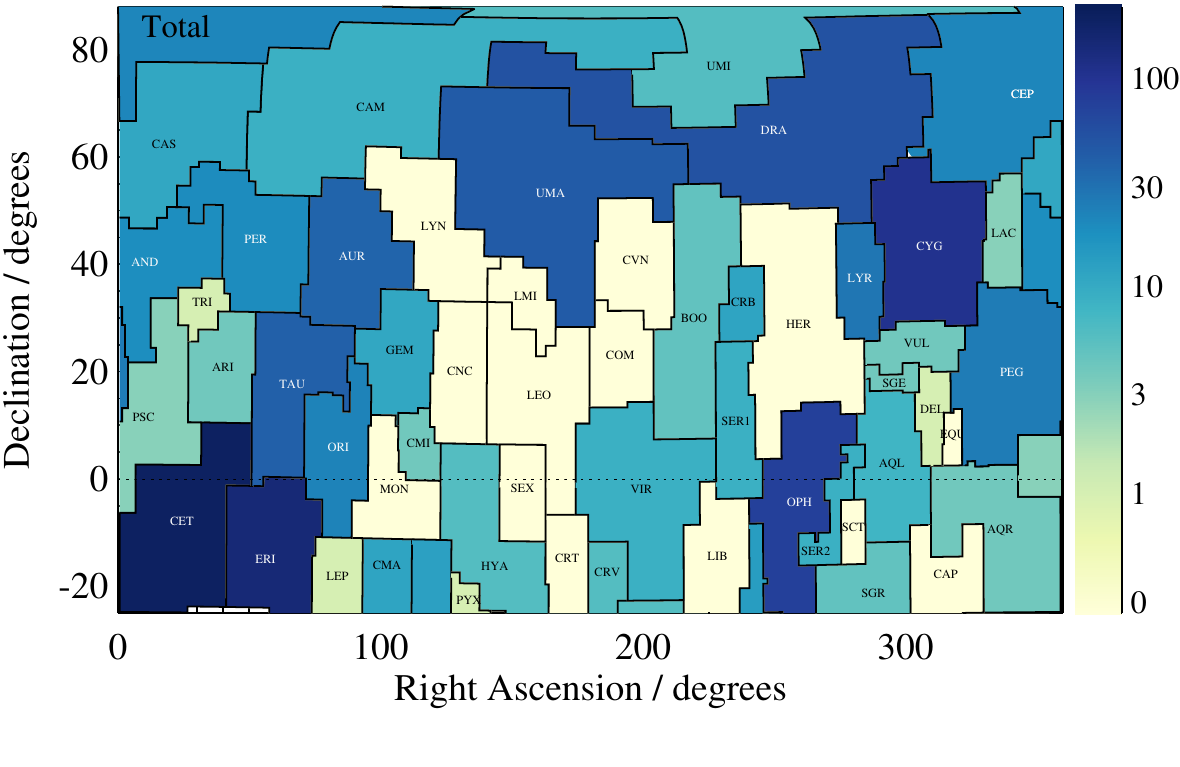}
\caption{Total references to constellation names by genitive form.}\label{fig:Totalgen}
\end{figure*}

Quantitative results are tabulated in Appendix B. In Figure \ref{fig:byname} we show heat maps of the frequency with which constellation names appear in each journal (or group of journals), with more frequently mentioned regions indicated by darker shades. For the purposes of visualisation, we use the constellation boundary data of Davenhall and Leggett (1989 \cite{1997yCat.6049....0D}) 
Despite some variation between magazine collections, the maps show some interesting similarities. The constellations Orion and Andromeda appear with high frequency in all sources considered. While zodaical constellations are generally more likely to be mentioned than others, a handful of outliers including Cygnus and Draco also recur in the distributions. The polar constellation Camelopardalis appears with very low frequency in all sources, as do the small constellations Lacerta, Equuleus and Delphinus. Leo Minor and Ursa Minor are also rarely mentioned, although in both cases this is likely due to a preference for their major constellation equivalents.

The distribution of mentions of constellation name in genitive form is presented in Figure \ref{fig:genitive}. This significantly changes the distribution of hits. Cyngi and Ceti now stand out as frequently mentioned across all sources considered here, with other genitives such as Eridani and Ophiuchi appearing with high frequency but varying between sources. 

Given the relatively minor variation between sources, we consider the summation of results for all magazine collections in Figures \ref{fig:Totalname} and \ref{fig:Totalgen}. Unsurprisingly, Andromeda and Orion remain the most commonly mentioned constellation names,  followed by Pegasus, Gemini, Taurus and Sagittarius. The genitive forms Ceti and Eridani are followed in appearance frequently by Cygni, Ophiuchi, Draconis and Aurigae.  These totals also reveal the least commonly invoked constellations: Lacerta, Scutum, Delphinus and Equuleus each appear a total of five times or less in either form across the archive collections considered, while Lepus, Pyxis,  Sextans and Vulpecula all appear fewer than ten times.

\section{Discussion}



\subsection{Considerations for Exoplanet and Astrobiology Searches}\label{sec:astrobio}

The consistency of constellation references across a range of science fictional sources spanning half a century of pulp magazine publication is interesting and suggests a high degree of reliability in the final results. 

While the narrative evidence considered here favours observations of fields lying within the constellations of Andromeda, Orion, Cetus, Eridanus and Cygnus, the design of exoplanet and astrobiology searches is likely to be influenced by other considerations. The structure of our own Milky Way, in particular, leads to constraints. Any solar system within 1 kpc of the Sun lies in our local space, and its potential habitability is most likely to be influenced by local star forming regions (with associated hard UV radiation or supernovae exercising a negative influence \cite[see e.g.][]{2018MNRAS.475.1829S}) or by the higher stellar densities found in the Milky Way's spiral arm structure. Fields which lie at very high galactic latitudes (e.g. pointing out of the plane of the Milky Way disk) might be presumed to intercept fewer potential solar systems.
For deeper and more sensitive surveys, able to detect more distant solar systems, the dense central structures of the Milky Way might lead to source confusion and high extinction for sightlines passing through the Galactic Centre

\begin{figure*}
\begin{minipage}{1.15\textwidth}
\hspace*{-4cm}
\includegraphics[width=0.55\textwidth]{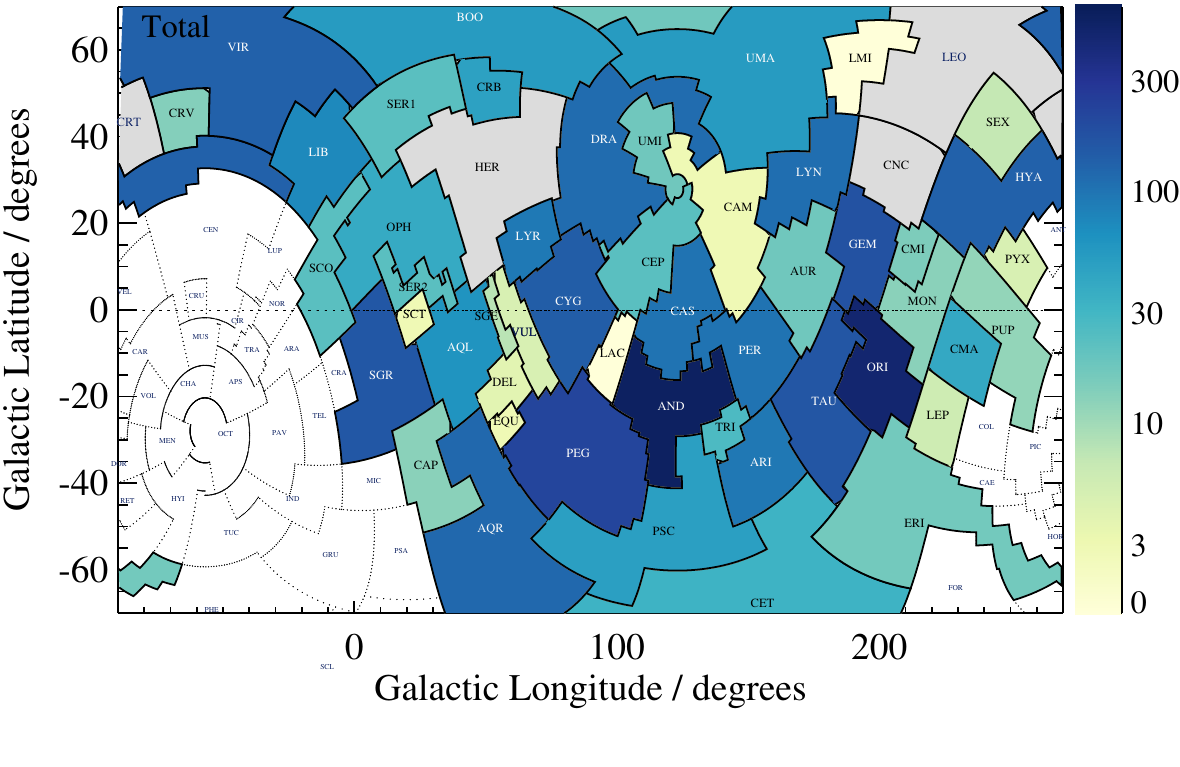}
\includegraphics[width=0.55\textwidth]{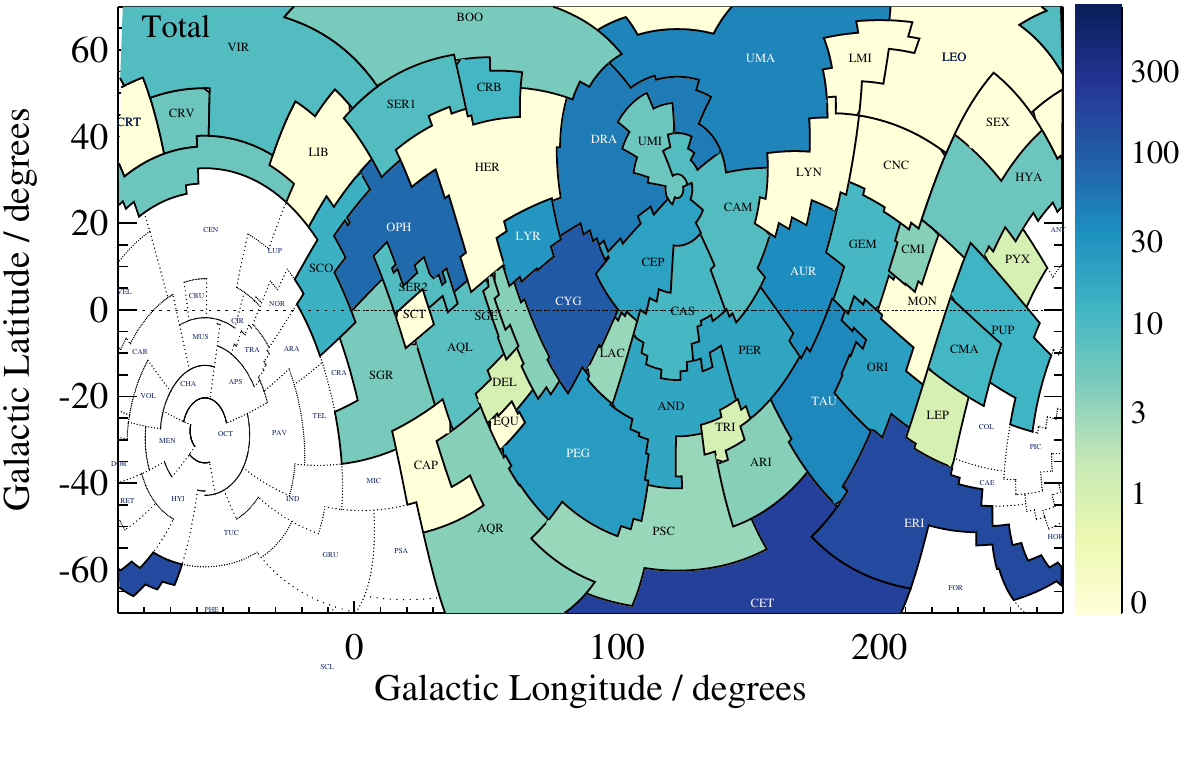}

\hspace*{-4cm}
\includegraphics[width=0.55\textwidth]{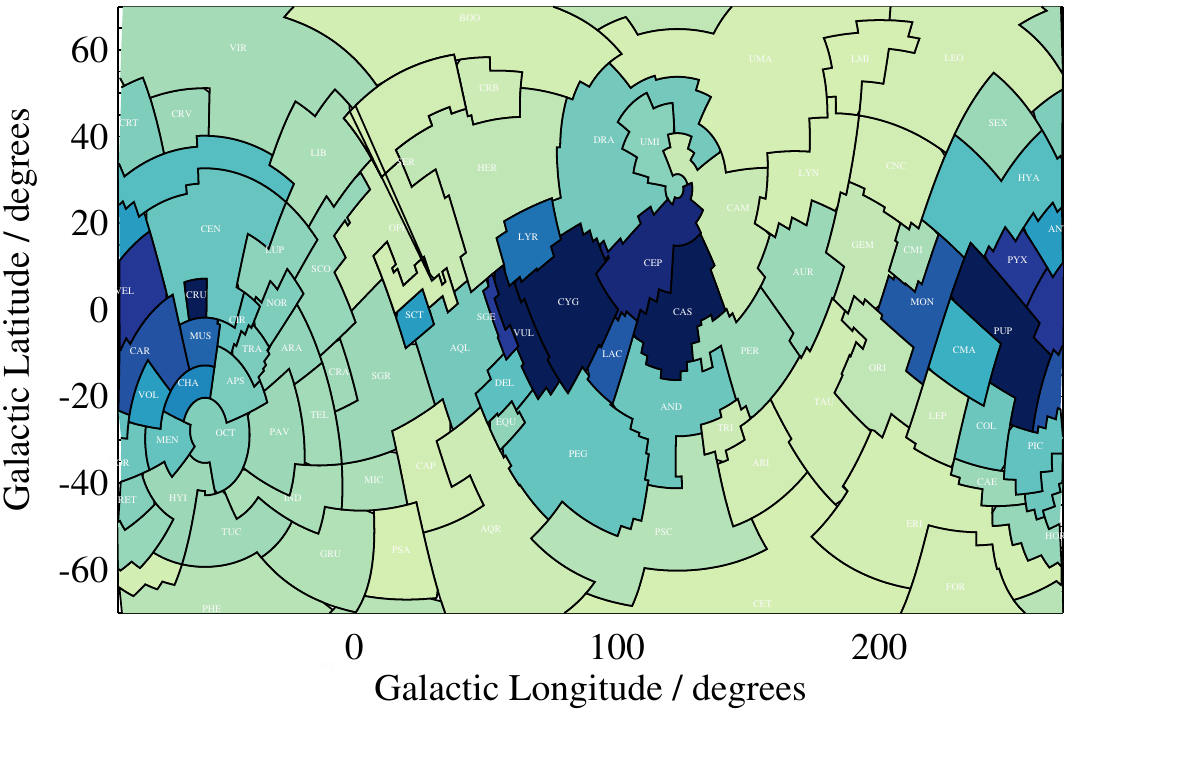}
\hspace*{1cm}\includegraphics[width=0.4\textwidth]{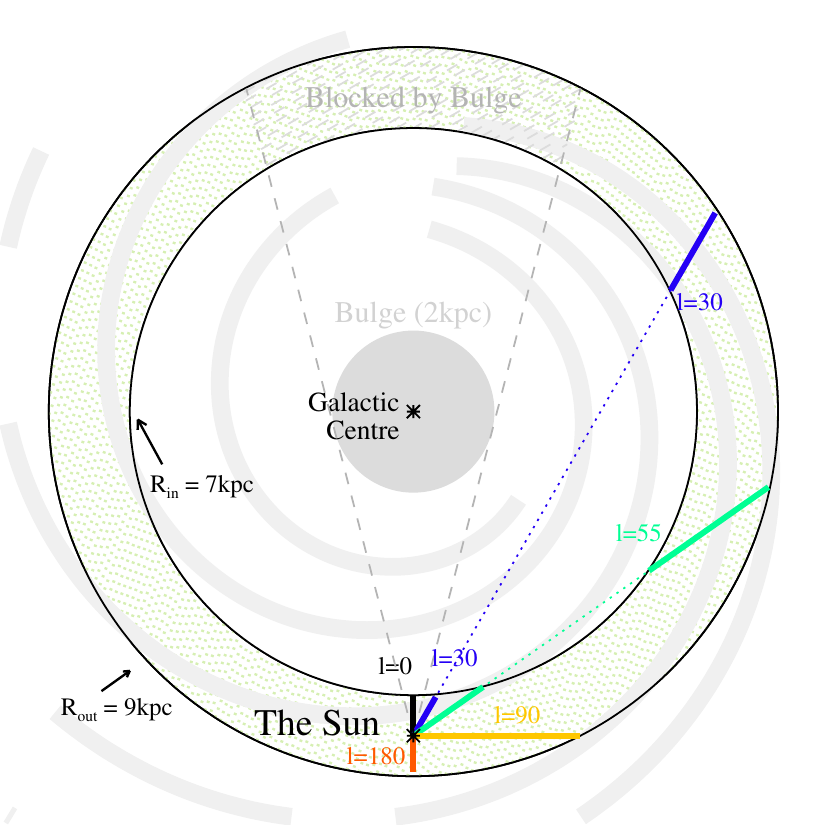}
\end{minipage}
\hspace*{-2cm}
\begin{minipage}{0.99\textwidth}
\caption{Density of SF references to constellations, shown in Galactic coordinates -- Top left: constellations by name; Top right: genitive form; Bottom left: habitable column (see section \ref{sec:astrobio}); Bottom right: model habitable zone, with approximate locations of spiral arms \cite{2014A&A...569A.125H} and the line of sight at a range of Galactic longitudes ($l$) indicated.}\label{fig:hab}
\end{minipage}
\end{figure*}

While the distribution of constellation mentions in SF shows little clear pattern when projected in right ascension and declination, given these considerations, it is interesting to consider whether there is any pattern relative to the Galactic plane. In Figure \ref{fig:hab} we recast the results of Figure \ref{fig:Totalname} and \ref{fig:Totalgen} in Galactic coordinates such that the Galactic Centre lies at (0,0) and the anti-centre at (180,0), with the plane of the Milky Way disk passing horizontally through the centre of the projection.

In the third panel of Figure \ref{fig:hab}, we compare this with an approximate estimate of the column density of habitable regions in the direction of each constellation. The Galactic habitable zone \cite{2004Sci...303...59L} has been identified as the region of the Milky Way disk located at at $7 < \mathrm{R / kpc} < 9$ from the Galactic Centre. Within this radius a combination of high stellar density and high metallicity may disfavour the long term survival of habitable exoplanets, while outside this radius low typical metallicities disfavour the formation of rocky planets. While this is a very simplistic model, neglecting the impact of stellar orbits and of spiral structures in the Galaxy, we adopt it here as a toy model. 

For the central coordinates of each constellation we calculate the projected path length in its direction that lies within the habitable zone. Lines of sight with an impact parameter of R$<2$\,kpc from the Galactic Centre are deemed to be blocked by the high extinction and stellar density of the Galactic bulge, such that only the foreground segment is counted. At some larger galactic longitudes, the line of sight may intersect the habitable zone twice when projected onto the plane of the disk. 

However for constellations centred at high galactic latitudes, these paths may be directed above or below the plane of the disk, particularly at large heliocentric radii. To account for this we weight line segments by $e^{-z/H}$, where $z$ is the projected height above the disk of the segment and $H=1$\,kpc is a conservative scale height for the thick disk. 

Unsurprisingly, the resultant map shows high column densities for constellations centred at low galactic latitudes and at longitudes of $70<b<130^\mathrm{o}$ or $250<b<310^\mathrm{o}$ that maximise the path length through the annular habitable zone. 

The overlap between regions with high potential habitability and those identified by the SF community is not striking. The region of interest around $l=270^o$ lies mostly in the southern hemisphere, and hence has been excluded from the study. However we note that Cygnus (and its genitive form cygni) represents a common region of high interest. It is also interesting to note that the constellations Lacerta and Scutum represent the inverse case: regions which lie in directions with a large habitable path length, but are seldom mentioned in speculative fiction.

\subsection{Dark Forests}

Given that we have already allowed the existence of precognitive impulses, it is logical to consider the influence of other potential mental abilities on the interpretation of our results. Were we to posit an advanced alien race with mental abilities, might they be able to affect (at a distance) the perception of their region by human beings? Making this assumption, three possibilities need to be considered. In the first, any hypothetical aliens are eager to make contact. In the second they are indifferent to contact. In either case, no modification would be expected to the predictive power of science fiction, except perhaps for a slight enhancement in its efficacy. The conclusions of the previous section are supported, or strengthened.

However an alternative possibility exists. The Dark Forest hypothesis is based on a game theory analysis (derived from \cite{fudenberg_1983_sequential}) which suggests that if there is any chance at all of an alien species in the Milky Way being hostile and technologically advanced (which cannot be ruled out), a species' best interests are served by opting to avoid contact at all costs, rather than making attempts to message extraterrestrial civilisations. This has been likened to the best strategy for a defenceless individual lost in a dark forest full of predators, notably by science fiction author Liu Cixin (2007), but also by earlier authors including Leinster (1945), Renmore (1976) and Bear (1987). 


Under this premise, the impact of remote mental manipulation may be to suppress thoughts of planets hosting advanced life: The constellations which are \textit{least} mentioned may be the most deserving of attention, if we want to locate the most philosophically and mentally advanced aliens. We conclude that the constellations \textit{Equuleus}, \textit{Lacerta}, \textit{Pyxis} and \textit{Scutum} should be closely studied for evidence of (well-concealed) bio- or technosignatures. Of these, Lacerta and Scutum also lie in the plane of the Milky Way disk, and so may represent the strongest targets.

\subsection{Caveats and Uncertainties}

There are clear limitations in our search methodology. It relies on indexing of scanned documents, much of it through optical character recognition and so subject to error and confusion. Manual removal of duplications is suboptimal, but necessary due to the absence of uniform metadata which can be used for automatic identification. We have also not distinguished here between science fact articles or science fictional accounts of human or alien activity. In each case, the implication is that members of the science fiction community anticipates future events to be influenced from the region, and thus implies the possibility of habitable planets (and/or exomoons or other locations).

\begin{figure}
\begin{center}
    
\includegraphics[width=0.85\textwidth]{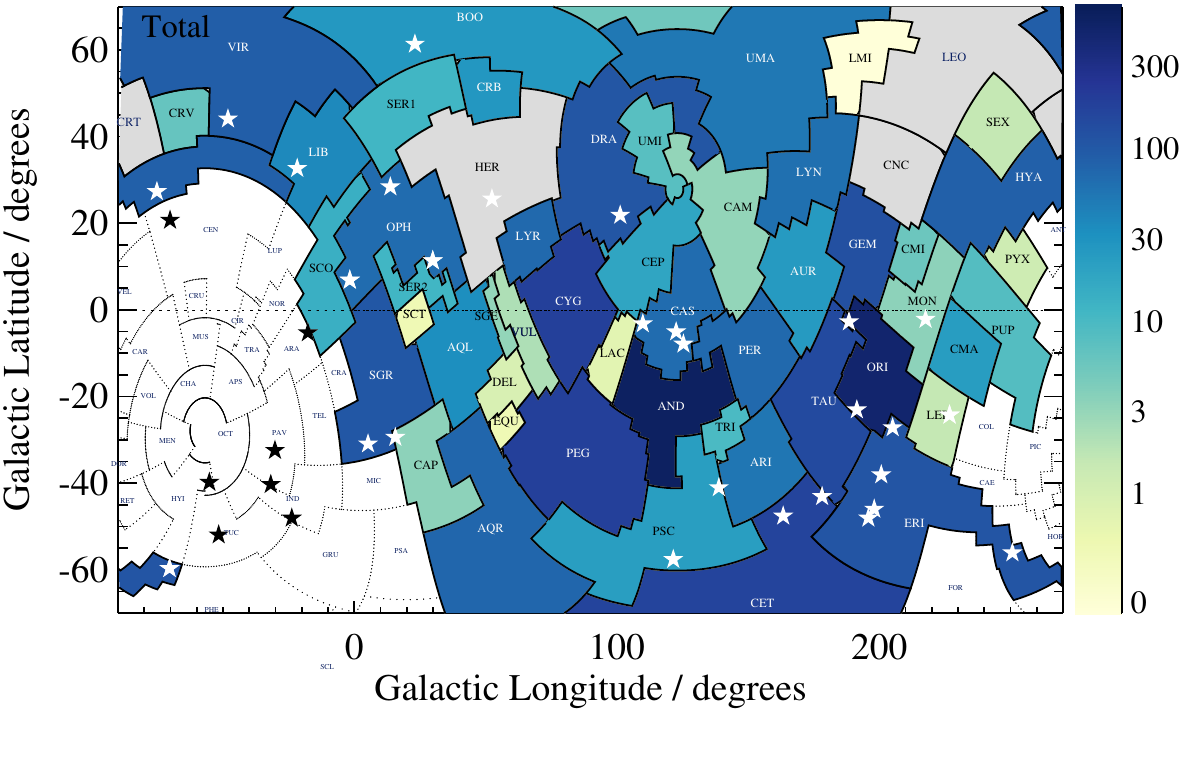}
\caption{The sum of results for constellation names and genitive forms across all magazines, with local ($d<10$\,pc) Sun-like (FGK) stars overplotted.}\label{fig:stars}
\end{center}
\end{figure}

We note that a subset of constellations appear far more prominently when the genitive form of their name is considered rather than the proper name. In this case, there are a handful of individual stars which may well be skewing the search results.  In Figure \ref{fig:stars} we overplot the total of name and genitive mentions across all magazines with the locations of local Sun-like stars. We select these from the Gaia Catalog of Nearby Stars \cite{2021A&A...649A...6G}, identifying sources with an estimated distance $<$10\,parsec from the Sun, and with a Gaia G-band absolute magnitude $3<M_G<7$ which corresponds to F, G and K classes for main sequence stars. There is no clear correspondence between the density of solar-type stars and the frequency of mentions in science fiction magazines, although a few exceptional cases do exist.

The constellation Cetus appears in science fiction primarily in its genitive form. A search for the phrase "Tau Ceti" confirms that in the majority of cases, these are references to this named star (accounting for 97 of the 121 search results). Tau Ceti has long been a popular target for science fiction authors. It is the closest, stable sun-like single star, just 12 light years from Earth, and is thus often treated as a prime candidate for both hosting habitable planets and for human colonisation. The system is now known to host at least four planets detected through the radial velocity method \cite{2017AJ....154..135F} and a debris disk \cite{2016ApJ...828..113M}.


Epsilon Eridani is another nearby star (10.5 light year) and accounts for 49 of the 90 references for the genitive form Eridani. It is an early K star, slightly smaller than Sol. The system is relatively young, and also hosts a debris disk \cite{1998ApJ...506L.133G}. There is also evidence for a Jupiter-like planet \cite{2000ApJ...544L.145H}, albeit in a closer and more elliptical orbit than Jupiter itself. 
In a science fictional context, the Eridanus constellation is also recognised as the location of the planet Vulcan in the television series \textit{Star Trek}, which canonically orbits 40 Eridani. More recently, 40 Eridani (as well as Tau Ceti) also appears in the blockbuster science fiction movie \textit{Project Hail Mary} (2026), based on a 2021 novel of the same name by Andy Weir.

Similarly, the 61 Cygni binary system accounts for 57 of the 104 Cygni references. This system is a close visual binary of two K-type stars around 11 light years from Earth. It attracted the interest of both astronomers and the science fiction community when variations in the binary orbit  were interpreted as possible evidence for tertiary companion in the form of a super-Jupiter planet in 1942 \cite{1943PASP...55...29S}. However these data were superseded by others which ruled out this interpretation. Despite increasingly sensitive observations, and repeated claims of detections, there has been no planet confirmed in orbit of the system.

A number of other constellations considered in this study host well-known  naked-eye stars. Examples include Vega in Lyra and Altair in Aquila. However in these cases the star name does not include the genitive and there is no clear evidence of it significantly skewing the search results. It should be noted that even if excluding the influence of the individual stars discussed above, a significant number of references to each constellation remains.

\section{Conclusions}

We have identified a number of constellations that have attracted, or avoided, interest in the science fiction community through the medium of the classic pulp (and digest) magazines. The most frequently mentioned constellations are Andromeda, Orion, Cetus and Eridanus. The least frequently mentioned constellations include Lacerta and Scutum. 

We conclude that science fiction pulp magazines yield a surprisingly consistent and clear indication of broadly defined regions where, in the opinion of the science-literate and engaged science fiction community, life may be found in the Milky Way - either now or in the future. Given the proven predictive power of these narrative evidence sources, we advise exoplanet search surveys to carefully consider the indicated regions in their survey design - either directly or under the Dark Forest hypothesis. 

\funding{This research received no funding whatsoever.}


\conflictsofinterest{The authors declare no known conflicts of interest.} 

\section*{Bibliography}

\bibliography{AprilUnum}


\newpage

\section*{Appendix A: Is there intelligent life somewhere out in space?\footnote{\textit{'Cause there's bugger all down here on Earth!}, (Monty Python, 1983)}}

This is, of course, a contribution in the long tradition of April Fools Day papers in astrophysics, and not intended to be taken seriously.  The methodological problems are obvious and many. However it's worth making a few more serious points here.

While science fiction is often credited with predictive power, it's important to stress that this is a result of logical extrapolation from extant knowledge and technologies, rather than precognitive abilities. In the peak era of the pulp magazines in the 1930s-60s, this included extrapolation of then-theorised or experimental rocket and nuclear technologies, whereas throughout its history, science fiction extrapolated from the constant growth of communications technology to produce apparent predictions of smart phones and today's interconnected online world. While these lay beyond the technology accessible to the authors, they were based on sound physical principles known at the time, and could be envisaged as a logical extension of then-extant trends.

So was such an extrapolation possible in the case of potential alien origins? The more careful of the pulp SF writers (who included professional physicists and astronomers) would have been aware of the relative impact of massive stars versus sun-like stars on habitability of their environs. They would also have been aware of the limitations of using the constellations as indications of alien origins. The constellations each cover a vast area of the sky from the perspective of modern telescopes, and their naked-eye visible members are not meaningful arrangements of stars in three dimensions, with apparent neighbours often very different in distance from Earth. As a result, the more careful writers would be likely to identify individual stars rather than constellations as the potentially habitable origins of life forms.  

Most importantly though, science fiction is \textbf{fiction}. For many of the pulp writers, the origins of their aliens would have been considered far less important than the nature of those aliens, how that nature can be used to reflect on the traits of humanity, and how they can be used to construct a compelling and entertaining narrative. For many such writers, star and constellation names are more likely to be used effectively as set-dressing, providing authority to the portrayal through use of familiar and plausible scientific names rather than as carefully-considered selections.

There are entire classes of planets now known (e.g. hot jupiters, sub-Neptunes, super-Earths) that were not even imagined in the twentieth century. The unpredictable impact of solar irradiation, stellar multiplicity and atmospheric composition on planetary (or extra-planetary) habitability is only now beginning to be understood. Even the best read of science fiction writers could not have reached a detailed awareness of these by extrapolation, rather than simple chance, and certainly could not have mapped them onto the vast range of stars present in each constellation in Earth's sky.

So, with regret, I have to conclude that the analysis presented in this paper is best treated as science fiction.

\vspace{12pt}

Caveat: Of course, if I'm wrong and we do actually discover aliens broadcasting from Cetus or Cygnus, or hiding in Lacerta, Equuleus or Scutum, then I retract the above and claim to make the first truly robust evidence-based prediction of where to look for life outside our solar system.

\vspace{10pt}

For further discussions of the relationship between science and science fiction some readers may be interested in my blog which posts new content every two weeks at \\\textbf{\href{https://www.warwick.ac.uk/CosmicStories}{https://www.warwick.ac.uk/CosmicStories}}

\newpage

\section*{Appendix B: Quantitative Data}

\vspace{20pt}

Table B1: Total mentions by Constellation

\vspace{10pt}

\begin{tabular}{|l|l|l|r|r|}
\hline
  \multicolumn{1}{|c|}{Acronym} &
  \multicolumn{1}{c|}{Name} &
  \multicolumn{1}{c|}{Genitive} &
  \multicolumn{2}{c|}{Totals}  \\
  \multicolumn{1}{|c|}{} &
  \multicolumn{1}{c|}{} &
  \multicolumn{1}{c|}{Name} &
  \multicolumn{1}{c|}{by name} &
  \multicolumn{1}{c|}{genitive} \\
\hline
  AND & Andromeda & Andromedae &  572 & 19\\
  AQR & Aquarius & Aquarii & 125 & 4\\
  AQL & Aquila & Aquilae &  62 & 8\\
  ARI & Aries & Arietis &  98 & 4\\
  AUR & Auriga & Aurigae &  18 & 38\\
  BOO & Bootes & Bootis &  54 & 5\\
  CAM & camelopardalis & Camelopardalis &  3 & 9\\
  CNC & Cancer & Cancri & 1728 & 4\\
  CVN & Canes Venatici & Canum Vanaticorum & 18 & 0\\
  CMA & Canis Major & Canis Majoris &  42 & 12\\
  CMI & Canis Minor & Canis Minoris &  15 & 4\\
  CAP & Capricornus & Capricorni &  13 & 0\\
  CAS & Cassiopeia & Cassiopeiae &  106 & 11\\
  CEP & Cepheus & Cephei &  23 & 22\\
  CET & Cetus & Ceti &  34 & 201\\
  COM & Coma Berenices & comae Berenices &  15 & 0\\
  CRB & Corona Borealis & Coronae Borealis &  48 & 12\\
  CRV & Corvus & Corvi &  14 & 6\\
  CRT & Crater & Crateris &  1683 & 0\\
  CYG & Cygnus & Cygni &  146 & 109\\
  DEL & Delphinus & Delphini &  4 & 1\\
  DRA & Draco & Draconis &  116 & 54\\
  EQU & Equuleus & Equulei & 3 & 0\\
  ERI & Eridanus & Eridani &  17 & 154\\
  GEM & Gemini & Geminorum &  179 & 11\\
  HER & Hercules & Herculis &  425 & 15\\
  HYA & Hydra & Hydrae &  138 & 6\\
  LAC & Lacerta & Lacertae &  1 & 3\\
  LEO & Leo & Leonis &  790 & 8\\
  LMI & Leo Minor & Leonis Minoris & 1 & 0\\
  LEP & Lepus & Leporis & 6 & 1\\
  LIB & Libra & Librae & 76 & 0\\
  LYN & Lynx & Lyncis & 113 & 0\\
  LYR & Lyra & Lyrae & 95 & 29\\
  MON & Monoceros & Moncerotis & 13 & 0\\
  OPH & Ophiuchus & Ophiuchi &  40 & 78\\
  ORI & Orion & Orionis &  496 & 23\\
\hline\end{tabular}

\newpage

Table B1: Total mentions by Constellation (continued)

\vspace{10pt}

\begin{tabular}{|l|l|l|r|r|}
\hline
  \multicolumn{1}{|c|}{Acronym} &
  \multicolumn{1}{c|}{Name} &
  \multicolumn{1}{c|}{Genitive} &
  \multicolumn{2}{c|}{Totals}  \\
  \multicolumn{1}{|c|}{} &
  \multicolumn{1}{c|}{} &
  \multicolumn{1}{c|}{Name} &
  \multicolumn{1}{c|}{by name} &
  \multicolumn{1}{c|}{genitive} \\
\hline
  PEG & Pegasus & Pegasi &  227 & 26\\
  PER & Perseus & Persei &  105 & 20\\
  PSC & Pisces & Piscium &  50 & 3\\
  PUP & Puppis & Puppis &  12 & 13\\
  PYX & Pyxis & Pyxidis &  5 & 1\\
  SGE & Sagitta & Sagittae & 8 & 4\\
  SGR & Sagittarius & Sagittarii &  160 & 5\\
  SCO & Scorpius & Scorpii &  23 & 14\\
  SCT & Scutum & Scuti & 3 & 0\\
  SER1 & Serpens & Serpentis &  23 & 9\\
  SEX & Sextans & Sextantis &  7 & 0\\
  TAU & Taurus & Tauri &  166 & 42\\
  TRI & Triangulum & Trianguli &  27 & 1\\
  UMA & Ursa Major & Ursae Majoris & 54 & 45\\
  UMI & Ursa Minor & Ursae Minoris &  18 & 6\\
  VIR & Virgo & Virginis &  137 & 9\\
  VUL & Vulpecula & Vulpeculae & 5 & 4\\
\hline\end{tabular}

\newpage

\hspace*{-4cm}Table B2: Mentions by Constellation and Magazine

\vspace{10pt}

\begin{minipage}{1.15\textwidth}
\hspace*{-4cm}
\begin{sideways}
\begin{tabular}{|l|r|r|r|r|r|r|r|r|r|r|r|r|r|r|}
\hline
  \multicolumn{1}{|c|}{Acronym} &
  \multicolumn{2}{c|}{Amazing Stories} &
  \multicolumn{2}{c|}{Astounding Stories} &
  \multicolumn{2}{c|}{Galaxy SF} &
  \multicolumn{2}{c|}{Fantast \& SF} &
  \multicolumn{2}{c|}{Azimov's} &
  \multicolumn{2}{c|}{Others} &
  \multicolumn{2}{c|}{Totals} \\
  \multicolumn{1}{|c|}{} &
  \multicolumn{1}{c|}{name} &
  \multicolumn{1}{c|}{gen} &
  \multicolumn{1}{c|}{name} &
  \multicolumn{1}{c|}{gen} &
  \multicolumn{1}{c|}{name} &
  \multicolumn{1}{c|}{gen} &
  \multicolumn{1}{c|}{name} &
  \multicolumn{1}{c|}{gen} &
  \multicolumn{1}{c|}{name} &
  \multicolumn{1}{c|}{gen} &
  \multicolumn{1}{c|}{name} &
  \multicolumn{1}{c|}{gen} &
  \multicolumn{1}{c|}{name} &
  \multicolumn{1}{c|}{gen} \\
\hline
  AND &  85 & 1 & 66 & 0 & 49 & 0 & 77 & 8 & 85 & 1 & 211 & 9 & 572 & 19\\
  AQR &  15 & 1 & 5 & 1 & 17 & 1 & 17 & 0 & 19 & 0 & 53 & 1 & 125 & 4\\
  AQL &  23 & 2 & 4 & 4 & 2 & 0 & 3 & 1 & 13 & 0 & 18 & 1 & 62 & 8\\
  ARI &  26 & 2 & 11 & 2 & 8 & 0 & 12 & 0 & 11 & 0 & 31 & 0 & 98 & 4\\
  AUR &  3 & 2 & 4 & 8 & 3 & 5 & 2 & 7 & 1 & 2 & 6 & 14 & 18 & 38\\
  BOO &  9 & 2 & 10 & 1 & 13 & 0 & 2 & 0 & 3 & 1 & 18 & 1 & 54 & 5\\
  CAM &  0 & 3 & 0 & 2 & 1 & 1 & 0 & 0 & 1 & 1 & 2 & 2 & 3 & 9\\
  CNC &  293 & 0 & 118 & 0 & 186 & 0 & 250 & 1 & 223 & 1 & 659 & 2 & 1728 & 4\\
  CVN &  1 & 0 & 4 & 0 & 5 & 0 & 2 & 0 & 0 & 0 & 7 & 0 & 18 & 0\\
  CMA &  14 & 2 & 7 & 2 & 4 & 1 & 5 & 1 & 2 & 2 & 11 & 4 & 42 & 12\\
  CMI &  7 & 0 & 1 & 0 & 2 & 1 & 2 & 0 & 0 & 1 & 4 & 2 & 15 & 4\\
  CAP &  4 & 0 & 0 & 0 & 3 & 0 & 1 & 0 & 1 & 0 & 5 & 0 & 13 & 0\\
  CAS &  18 & 3 & 11 & 0 & 13 & 1 & 16 & 2 & 10 & 1 & 39 & 4 & 106 & 11\\
  CEP &  4 & 3 & 6 & 3 & 3 & 2 & 3 & 5 & 1 & 1 & 7 & 8 & 23 & 22\\
  CET &  2 & 18 & 3 & 23 & 4 & 18 & 6 & 26 & 5 & 36 & 15 & 80 & 34 & 201\\
  COM &  1 & 0 & 3 & 0 & 4 & 0 & 1 & 0 & 1 & 0 & 6 & 0 & 15 & 0\\
  CRB &  19 & 0 & 6 & 2 & 3 & 1 & 4 & 0 & 5 & 4 & 12 & 5 & 48 & 12\\
  CRV &  1 & 0 & 0 & 0 & 2 & 1 & 0 & 1 & 5 & 1 & 7 & 3 & 14 & 6\\
  CRT &  417 & 0 & 225 & 0 & 170 & 0 & 155 & 0 & 196 & 0 & 521 & 0 & 1683 & 0\\
  CYG & 29 & 16 & 20 & 29 & 17 & 13 & 24 & 15 & 8 & 4 & 49 & 32 & 146 & 109\\
  DEL & 2 & 1 & 1 & 0 & 1 & 0 & 0 & 0 & 0 & 0 & 1 & 0 & 4 & 1\\
  DRA & 14 & 11 & 11 & 3 & 11 & 11 & 17 & 0 & 18 & 9 & 46 & 20 & 116 & 54\\
  EQU & 0 & 0 & 0 & 0 & 1 & 0 & 1 & 0 & 0 & 0 & 2 & 0 & 3 & 0\\
  ERI & 4 & 11 & 0 & 15 & 3 & 11 & 4 & 21 & 0 & 32 & 7 & 64 & 17 & 154\\
  GEM & 25 & 0 & 7 & 1 & 19 & 5 & 32 & 0 & 23 & 0 & 74 & 5 & 179 & 11\\
 HER & 101 & 3 & 71 & 4 & 22 & 1 & 60 & 3 & 45 & 0 & 127 & 4 & 425 & 15\\
  HYA & 27 & 4 & 16 & 0 & 11 & 0 & 24 & 1 & 13 & 0 & 48 & 1 & 138 & 6\\
  LAC &  0 & 1 & 0 & 0 & 1 & 1 & 0 & 0 & 0 & 0 & 1 & 1 & 1 & 3\\
  LEO &  214 & 2 & 77 & 0 & 42 & 0 & 119 & 0 & 89 & 3 & 250 & 3 & 790 & 8\\
  \hline\end{tabular}
  \end{sideways}
  \end{minipage}
  
\newpage

\hspace*{-4cm}Table B2: Mentions by Constellation and Magazine (continued)

\vspace{10pt}

\begin{minipage}{1.15\textwidth}
\hspace*{-4cm}
\begin{sideways}
\begin{tabular}{|l|r|r|r|r|r|r|r|r|r|r|r|r|r|r|}
\hline
  \multicolumn{1}{|c|}{Acronym} &
  \multicolumn{2}{c|}{Amazing Stories} &
  \multicolumn{2}{c|}{Astounding Stories} &
  \multicolumn{2}{c|}{Galaxy SF} &
  \multicolumn{2}{c|}{Fantast \& SF} &
  \multicolumn{2}{c|}{Azimov's} &
  \multicolumn{2}{c|}{Others} &
  \multicolumn{2}{c|}{Totals} \\
  \multicolumn{1}{|c|}{} &
  \multicolumn{1}{c|}{name} &
  \multicolumn{1}{c|}{gen} &
  \multicolumn{1}{c|}{name} &
  \multicolumn{1}{c|}{gen} &
  \multicolumn{1}{c|}{name} &
  \multicolumn{1}{c|}{gen} &
  \multicolumn{1}{c|}{name} &
  \multicolumn{1}{c|}{gen} &
  \multicolumn{1}{c|}{name} &
  \multicolumn{1}{c|}{gen} &
  \multicolumn{1}{c|}{name} &
  \multicolumn{1}{c|}{gen} &
  \multicolumn{1}{c|}{name} &
  \multicolumn{1}{c|}{gen} \\
\hline
  LMI &  0 & 0 & 0 & 0 & 1 & 0 & 0 & 0 & 0 & 0 & 1 & 0 & 1 & 0\\
  LEP &  4 & 1 & 1 & 0 & 1 & 0 & 0 & 0 & 0 & 0 & 1 & 0 & 6 & 1\\
  LIB &  14 & 0 & 5 & 0 & 9 & 0 & 11 & 0 & 9 & 0 & 29 & 0 & 76 & 0\\
  LYN &  26 & 0 & 6 & 0 & 5 & 0 & 14 & 0 & 22 & 0 & 41 & 0 & 113 & 0\\
  LYR & 23 & 7 & 11 & 2 & 9 & 2 & 15 & 7 & 7 & 1 & 31 & 10 & 95 & 29\\
  MON &  0 & 0 & 0 & 0 & 4 & 0 & 1 & 0 & 2 & 0 & 7 & 0 & 13 & 0\\
  OPH & 12 & 6 & 7 & 6 & 7 & 11 & 0 & 12 & 4 & 10 & 11 & 33 & 40 & 78\\
  ORI & 91 & 6 & 48 & 3 & 30 & 3 & 78 & 3 & 71 & 1 & 179 & 7 & 496 & 23\\
  PEG &  27 & 3 & 13 & 1 & 14 & 3 & 56 & 1 & 24 & 7 & 94 & 11 & 227 & 26\\
  PER & 15 & 4 & 11 & 4 & 7 & 1 & 19 & 4 & 14 & 1 & 40 & 6 & 105 & 20\\
  PSC &  9 & 1 & 2 & 0 & 8 & 0 & 8 & 1 & 4 & 0 & 20 & 1 & 50 & 3\\
  PUP & 1 & 1 & 2 & 2 & 5 & 5 & 0 & 0 & 0 & 0 & 5 & 5 & 12 & 13\\
  PYX &  0 & 0 & 0 & 1 & 3 & 0 & 0 & 0 & 0 & 0 & 3 & 0 & 5 & 1\\
  SGE &  3 & 0 & 2 & 2 & 2 & 0 & 0 & 1 & 0 & 0 & 2 & 1 & 8 & 4\\
  SGR &  30 & 1 & 13 & 0 & 18 & 1 & 21 & 1 & 20 & 0 & 59 & 2 & 160 & 5\\
  SCO &  1 & 3 & 3 & 5 & 3 & 1 & 4 & 2 & 3 & 0 & 10 & 3 & 23 & 14\\
  SCT &  0 & 0 & 0 & 0 & 1 & 0 & 0 & 0 & 1 & 0 & 2 & 0 & 3 & 0\\
  SER &  1 & 3 & 3 & 2 & 2 & 1 & 5 & 1 & 3 & 0 & 10 & 2 & 23 & 9\\
  SEX & 1 & 0 & 3 & 0 & 2 & 0 & 0 & 0 & 0 & 0 & 2 & 0 & 7 & 0\\
  TAU &  27 & 11 & 16 & 9 & 17 & 4 & 25 & 5 & 20 & 2 & 62 & 11 & 166 & 42\\
  TRI & 1 & 1 & 3 & 0 & 3 & 0 & 2 & 0 & 7 & 0 & 12 & 0 & 27 & 1\\
  UMA & 13 & 2 & 11 & 1 & 8 & 0 & 13 & 2 & 9 & 19 & 1 & 21 & 54 & 45\\
  UMI &  4 & 0 & 1 & 0 & 1 & 2 & 4 & 0 & 2 & 1 & 7 & 3 & 18 & 6\\
  VIR & 24 & 1 & 14 & 0 & 14 & 2 & 25 & 0 & 11 & 2 & 50 & 4 & 137 & 9\\
  VUL &  0 & 0 & 0 & 0 & 2 & 2 & 1 & 0 & 0 & 0 & 3 & 2 & 5 & 4\\
\hline\end{tabular}
\end{sideways}
\end{minipage}

\end{document}